\documentclass[Times,12pt,oneside,openany,print,index]{report}
\usepackage[a4paper,width=150mm,top=25mm,bottom=25mm]{geometry}
\usepackage[english]{babel}
\usepackage[utf8]{inputenc}

\usepackage{listings}
\usepackage{csquotes} 
\usepackage{amsmath} 
\usepackage[toc,page]{appendix}

\usepackage{graphicx} 
\graphicspath{/images} 
\usepackage{caption} 
\usepackage{array} 

\usepackage[nottoc]{tocbibind} 

\usepackage[normalem]{ulem} 

\usepackage{hyperref} 
\hypersetup{
    colorlinks=true,
    linkcolor=blue,
    filecolor=magenta,      
    urlcolor=blue,
    citecolor=blue,
}

\usepackage{amsmath,amssymb,amsfonts}
\usepackage{algorithmic}
\usepackage{graphicx}
\usepackage{textcomp}
\usepackage[table,xcdraw]{xcolor}
\usepackage{float}
\usepackage{longtable}
\usepackage{supertabular}

\usepackage{nomencl} 
\renewcommand{\nompreamble}{The next list describes several symbols \& abbreviation that will be later used within the body of the document}
\makenomenclature

\usepackage[backend=biber,style=ieee,sorting=ynt]{biblatex} 
\let\cleardoublepage=\clearpage 

\usepackage{multirow}
\usepackage{float}

\DeclareUnicodeCharacter{1EF3}{\`y}

\begin{document}

\thispagestyle{empty} 
\begin{titlepage}
\renewcommand*{\thepage}{Title} 

    \begin{center} 
        \vspace*{3cm} 
        
        {\fontsize{16pt}{22pt}\selectfont{Combining Machine Learning Classifiers for Stock Trading with Effective Feature Extraction }
        } 
        \vspace{1.5cm}
        
        \text{by}
        
        \vspace{0.5cm}
        
        	A. K. M. Amanat Ullah\\
	        amanat.ndc@gmail.com\\
	          \vspace{0.5cm}
	        Miftah Uddin Md Ihsan\\
	       miftah.uddin.mohammad.ihsan@g.bracu.ac.bd\\
	         \vspace{0.5cm}
	        Fahim Imtiaz\\
	        fahim.imtiaz@g.bracu.ac.bd\\
	         \vspace{0.5cm}
	        Md. Golam Rabiul Alam\\
	        rabiul.alam@bracu.ac.bd\\
	        \vspace{0.5cm}
	        Mahbub Majumdar\\
	       majumdar@bracu.ac.bd\\
 \vspace{2.5cm}
        
    		Department of Computer Science and Engineering\\
            Brac University\\
 \vspace{2.5cm}
 \textbf{Reference} to this paper should be made as follows:\\
 Ullah, A.K.M.A., Imtiaz, F., Ihsan, M.U.M.,
Alam, M.G.R. and Majumdar, M. (2023) ‘Combining machine learning and effective feature
selection for real-time stock trading invariable time-frames’, Int. J. Computational Science and
Engineering, Vol. 26, No. 1, pp 28-44 DOI: \url{https://doi.org/10.1504/IJCSE.2023.129152}

    \end{center}

\end{titlepage} 
\cleardoublepage

\pagenumbering{roman} 




\phantomsection
\addcontentsline{toc}{chapter}{Abstract}
\section*{Abstract}

The unpredictability and volatility of the stock market render it challenging to make a substantial profit using any generalized scheme. Many previous studies tried different techniques to build a machine learning model, which can make a significant profit in the US stock market by performing live trading. However, very few studies have focused on the importance of finding the best features for a particular period for trading. Our top approach used the performance to narrow down the features from a total of 148 to about 30. Furthermore, the top 25 features were dynamically selected before each time training our machine learning model. It uses ensemble learning with four classifiers: Gaussian Naive Bayes, Decision Tree, Logistic Regression with L1 regularization and Stochastic Gradient Descent, to decide whether to go long or short on a particular stock. Our best model performed daily trade between July 2011 and January 2019, generating 54.35\% profit. Finally, our work showcased that mixtures of weighted classifiers perform better than any individual predictor about making trading decisions in the stock market.

\vspace{1cm}
\textbf{Keywords:}  Feature Selection, Feature Extraction, Stock trading, Ensemble learning
\pagebreak





\cleardoublepage

\pagenumbering{arabic} 

\chapter{Introduction}

\section{Background} 
Stocks are essentially small pieces of ownership of a company, and the stock market works like an auction where investors buy and sell stocks. Owning stock means the shareholder owns a proportion of the company equal to the number of shares the person bought, against the total outstanding shares of the company. For example, if a company has 1 million shares and an individual owns 50,000 shares of the company, the person has a 5\% stake in it.

\subsection{Long-short investment strategy}

Traditionally, stock investing was focused on looking for stocks to buy long that is likely to appreciate \cite{jacobs1993long}. There was little, if any, thought given to capitalizing on short-selling overvalued stocks. When investors began to employ both long and short strategies in their investment portfolio more benefits and opportunities presented themselves which was previously unavailable.

Buying long is simply buying a stock that you think will appreciate, and selling for profit when the stock price rises. For instance, imagine that you bought 500 shares of a particular stock, at \$10 per share. This amounts to \$5000. After a week, the price of a share of ABC rises to \$55. You sell the stock, pocketing a profit of \$500.

Shorting is when you borrow stocks that you expect will depreciate from a broker, at interest, and selling them while you wait for the price to drop. Once the price has lowered a significant amount, you pay back the lender by buying the same number of stocks that you borrowed in the first place, at the lower price. Your profit is the difference in price minus the interest and commissions.

For instance, you borrow 100 shares of XYZ, at \$50 per share, and immediately sell them for \$5000 while waiting for the share price to depreciate. Once the price per share of XYZ has dropped to \$45, you buy 100 shares of XYZ and pay \$4500 for it. Return the 100 shares to the lender and whatever remains minus the interest and commissions is the profit. In this case, your profit is \$500.

\section{Motivation}
\textbf{"I will tell you how to become rich. Close the doors. Be fearful when others are greedy. Be greedy when others are fearful." — Warren Buffett.} The quote suggests that trading decisions need to be made entirely based on logic and not based on human emotions. Oftentimes people cannot control their emotions It is difficult to let out emotion while trading. Effective trading involves making decisions without letting emotions getting in the way. The perfect way to solve this problem is to deploy a machine that solely relies on logic to make effective decisions. On another note, current estimates show that automated trading accounts for 50-70 percent of equities trades in the United States, 40\% in Canada, and 35\% in London \cite{o2012high, grant2011high}. Therefore there will come a time where all the trades will be managed by machines. To prepare the world for such a time more research into this field is vital. We are all aware of the unpredictability of the stock market, and how difficult it is to predict because of the noise in the data \cite{doi:10.1080/09540091.2015.1039492}. Some people believe that it is not possible to do so. We believe that with the advancements in Machine Learning algorithms and Artificial Intelligence, we can predict stock market trends sufficiently, given we provide sufficient, refined, data to our models. Many previous researches \cite{yuan2020integrated,tsai2010combining,he2013feature} worked with selecting features with different algorithms for maximizing profit. However most of them did not run the final test on an actual stock trading setting. Additionally, none of the research worked on which feature time-frame works best for how many days of trading. Our research aims to explore this research gap.

 \section{Contributions}
 To our knowledge, Our model is the first to recommend which feature time-frame suitable for how many days of trading. To address this problem our novel approach calculated each of the features using different time-variants (default, 1day, 2 days, 5 days, 22 days) to find out which variant works best for daily trading, weekly trading and monthly trading. Using the recommended features our model further used dynamic feature selection techniques coupled with advanced machine learning algorithms to generate profit on real-time stock data from 1500 stocks from the US stock market from 2011 to 2019 and generated significant profit which is on par and in some cases better than the state-of-the-art models. The main contributions of this research are as follows:
 \begin{itemize}
 \item  A Dynamic Feature Selection mechanism has been proposed to select discriminative features over multiple time-frames for holding long and short positions for effective stock trading.
 
 \item After the initial feature selection mechanism the proposed model uses ANOVA for finalizing the set of features and uses ensemble of various machine learning algorithms for stock trading that generated 54.35\% profit on the initial investment.

 \end{itemize}

 \section{Paper Organization}
 In the next section of our paper, we review the literature of the previous work on machine learning models in order to predict trends in the stock market. We discuss our overall strategy in chapter 3. The dataset analysis is discussed in Chapter 4. Chapter 6 is on the result analysis. Chapter 5 talks  The final chapter concludes the paper and talks about the limitations in our thesis and future prospects.

\chapter{Literature Review}
The prevalence of volatility in the stock market and also other markets (e.g. Forex\cite{tiong2016forex} and Cryptocurrency \cite{chen2020dependence}) makes predicting stock prices anything but simple. Before investing, investors perform two kinds of analysis \cite{patel2015predicting}. The first of these is fundamental analysis, where investors look into the value of stocks, the industry performance, economic factors, sentiment analysis \cite{chang2020sentiments} etc. and decide whether or not to invest. Technical analysis is the second, more advanced, analysis that involves evaluating those stocks through the use of statistics and activity in the current market, such as volume traded and previous price levels \cite{patel2015predicting}. Technical analysts use charts to recognize patterns and try to predict how a stock price will change. Malkiel and Fama’s Efficient market hypothesis states that predicting the values of stocks considering financial information is possible because the prices are informationally efficient \cite{malkiel1970efficient}. As many unpredictable variables influence stocks and the stock market in general, it seems logical that factors such as the public image of the company and the political scenario of a country will be reflected in the prices. By sufficiently preprocessing the data obtained from stock prices and the algorithms and their factors are appropriate, it may be possible to predict stock or stock price index.

There were quite a few different implementations of machine learning algorithms for the purposes of making stock market price predictions. Different papers experimented with different machine learning algorithms that they implemented in order to figure out which models produced the best results. Dai and et al. attempted to narrow down the environment by selecting certain criteria \cite{Dai_automatedstock}. Under these criteria, they were able to achieve a profit of 0.0123, recall 30.05\%, with an accuracy of 38.39\%, and 55.07\% precision, using a logistic regression model, after training the model for an hour. Zheng and Jin observed that when compared with Logistic Regression, Bayesian Network, and a Simple Neural Network, a Support Vector Machine having radial kernel gave them the most satisfactory results  \cite{zheng2017using}. Due to their limited processing power, they were only able to use a subset of their data for training their model and recommended that a more powerful processor be used to achieve better results. Similar recommendations were made by G. Chen and et al., stating that their preferred model, the Long Short-Term Memory (LSTM) \cite{doi:10.1080/09540091.2021.1940101}, would have performed better were they able to train the different layers and neurons using higher computing power \cite{chenapplication}. Since the data was non-linear in nature, a Recurrent Neural Network (RNN) would be more suited to the task. Recent researches also showcase the use of Transformer Networks \cite{hu2021local} and fuzzification \cite{hu2021local} techniques in stock trading and prediction.

In \cite{hegazy2014machine} it was discussed that when performing stock price prediction, it came out to be that ANN the algorithm that was once popular for prediction suffers from overfitting due to large numbers of parameters that it needs to fix \cite{tao2004input}. This is where support vector machine (SVM) came into play and,, it was suggested, that this method could be used as an alternative to avoid such limitations, where according to the VC theory \cite{vapnik2013nature} SVM calculates globally obtained sol unlike the ones obtained through ANN which mostly tend to fall in the local minima. It was seen that using an SVM model the accuracy of the predicted output came out to be around 57\% \cite{kim2003financial}. There is one other form of SVM and that is LS-SVM (Least squared support vector machine). In the paper \cite{madge2015predicting} it was mentioned that if the input parameters of LS-SVM is tuned and refined then the output of this classification algorithm boosts even further and shows promise to be a very powerful method to keep an eye out for. SVM being this powerful and popular as is it, is now almost always taken into consideration when it comes to predicting price of a volatile market, and thus we think that incorporating this into our research will boost our chances of getting a positive result.

While classical regression was more commonly used back in the day, non-linear machine learning algorithms are also increasingly being used as trading data regarded as time-series data which is non-stationary in nature. However, Artificial Neural Networks and SVM remain among the most popular methods used today. Every algorithm has a unique learning process. ANN simulates the workings of the human brain by creating a network of neurons \cite{patel2015predicting}. The Hidden Markov Model (HMM), Artificial Neural Networks (ANN) as well as Genetic Algorithms (GA) were combined into one fusion model in order to predict market behavior \cite{hassan2007fusion}. The stock prices converted to distinct value sets using ANN, which then became the input for the HMM. Using a selection of features determined from ARIMA analyses, Wang and Leu \cite{wang1996stock} designed a prediction model  which was helpful in predicting market trends in the Taiwanese stock market. This produced an acceptable level of accuracy in predicting market trends of up to 6 weeks, after the networks were trained using 4-year weekly data \cite{patel2015predicting}. A hybridized soft computing algorithm was defined by Abraham and et al. for automatic market predictions and pattern analysis \cite{abraham2001hybrid}. They made used the Nasdaq-100 index of the Nasdaq stock market for forecasting a day ahead with neural networks. A neuro-fuzzy system was used to analyze the predicted values. This system produced promising results. A PNN (probabilistic neural network) model was trained using historical data by Chen and et al. for investment purposes \cite{chen2003application}. When set against other investment strategies, namely the buy and hold concept and those which made use of forecasts estimated by the random walk and parametric Gaussian Mixture Model, PNN-based investment strategies produced better results.

By searching a higher dimension hyperplane, a well-known SVM algorithm that separates classes was developed by Vapnik \cite{vapnik1999overview}. To test the predictability of price trends in the NIKKEI 255 index, Wang and et al. used a SVM to make forecasts \cite{huang2005forecasting}. They also made comparisons with other methods of classification, such as Elman Backpropagation Neural Networks, Quadratic Discriminant Analysis, and  Linear Discriminant Analysis, SVM produced better experimental results. Kim compared the use of SVM to predict the daily stock price direction against Case-Based Reasoning and neural network in the Korean stock market \cite{kim2003financial}. The initial attributes were made up of twelve technical indicators. SVM was proven to have produced better results.

Ensemble methods such as random forests help to reduce the probability of the data overfitting. Random forests use decision trees and majority voting to obtain reliable results. In order to perform an analysis on stock returns, Lin and  et al. tested a prediction model that used the classifier ensemble method  \cite{tsai2011predicting} and took bagging and majority voting methods into consideration. It was found that models using single classifiers under-performed compared to the ones using multiple classifiers, in regards to ROI and accuracy when the performances of those using an ensemble of several classifiers and those using single baseline classifiers were compared \cite{patel2015predicting}. An SVM ensemble-based Financial Distress Prediction (FDP) was a new method proposed by Sun and Li \cite{sun2012financial}. Both individual performance and diversity analysis were used in selecting the base classifiers from potential candidates for the SVM ensemble. The SVM ensemble produced superior results when compared to the individual SVM classifier. A sum of ten data mining techniques, some of which included KNN, Naive Bayes using kernel estimation, Linear Discriminant Analysis (LDA), Least Squared SVM, was used by Ou and Wang to try and forecast price fluctuations in the stock market of Hong Kong  \cite{ou2009prediction}. The SVM and LS-SVM were shown to produce better predictions compared to the other models.

\section{Background Study of Feature Selection in stock trading}

The study \cite{he2013feature} measured twelve technical indicators for further investigation using data from the Shanghai Stock Exchange Composite Index (SSECI) from March 24, 1997 to August 23, 2006. The stock market's input variables were chosen from a total of 12 indicators. SMA, EMA, ALF (Alexander's filter), Relative Strength, RSI, MFI, percent B Indicator, Volatility, Volatility Band, CHO (Chaikin Oscillator), MACD (Moving Average Convergence-Divergence), percent K Indicator, Accumulation and distribution (AD) oscillator, and Williams percent R indicator are some of the indicators used. Then, PCA (Principal Component Analysis), Genetic Algorithm, and Sequential Forward Feature Selection methods to select which features for optimal investment. However, The paper did not include any resulting analysis or graphical representations of the results.

Yuan and et al. \cite{yuan2020integrated} selected 60 features for their prediction. The data comes from the Chinese A-share market and dates from January 1, 2010 to January 1, 2018. The algorithms used for prediction were Support Vector Machine(SVM), Artificial Neural Networks(ANN) and Random Forest. For the Feature selection, the paper used Recursive Feature Elimination (RFE) and Random Forest Feature selection using the information gain values. The Random Forest(RF) for feature selection and RF model for prediction has the greatest annualized return when it picks the top 1\% of companies, with a 29.51 percent annualized return. The RF-RF model's profitability is further investigated using the stratified back-testing technique, and the new long-short portfolio's annualized return from 2011 to 2018 is 21.92 percent, with a maximum drawdown of just 13.58 percent. This profit is not substantial for proving the success of their model because better results can be achieved.

To decrease the cost of training time and increase prediction accuracies, the work \cite{huang2009hybrid} of Hunag and et al. combined the Support Vector Regressor (SVR) with the self-organizing feature map (SOFM) method and a filter-based feature selection. Thirteen technical indicators were used as input variables to forecast the daily price in the Taiwan index futures (FITX) in order to forecast the price index for the next day. The SOFM-SVR with feature selection had a Mean Absolute Percentage Error (MAPE) of 1.7726 percent, which is higher than the single SVR with feature selection and the one without feature selection. However, They did not test their strategy in the real stock market which would further evaluate their model's actual performance. 

Barak and et al proposed a hybrid feature selection method \cite{barak2015wrapper} using ANFIS (Adaptive Neural Fuzzy Inference System) and the ICA (Imperialist Competitive Algorithm) is used to choose the most suitable features. The trading signals generated by the model achieved superior outcomes with 87 percent prediction accuracy, and the wrapper features selection achieves a 12 percent increase in predictive performance over the basic research. Furthermore, since wrapper-based feature selection models are much more time-consuming, the results of our wrapper ANFIS-ICA method are better in terms of reducing time and improving prediction accuracy when compared to other algorithms like the wrapper Genetic algorithm (GA). However, they worked on only 24 features at max and did not test implement a long-short strategy.

The research \cite{nti2019random} by Nti and et al. used Random Forest (RF) with an improved leave-one-out cross-validation strategy and a Long Short-Term Memory (LSTM) Network to evaluate the degree of importance between various sectors stock-price and MVs and forecasted a 30-day had stock-price. From January 2002 to December 2018, the research dataset was acquired from the GSE official website, and the 42 macroeconomic indicators dataset was collected from the Bank of Ghana (BoG) official website. The LSTM model performed better than the baseline ARIMA model. But real-time trading was not done in this study. 

The paper \cite{gandhmal2021wrapper} used 12 technical features. The features are selected using decision tree algorithm based on wrapper feature selection. The paper uses the chronological penguin Levenberg–Marquardt-based nonlinear autoregressive network (CPLM-based NARX) for prediction. The suggested paper showed that CPLM-based NARX outperformed the competition in terms of MAPE and RMSE, with values of 0.96 and 0.805, respectively in comparison with the Regression model, Deep Belief Network (DBN), and NeuroFuzzy-Neural Network. This study does not analyze between the different timeframes of each technical feature and only uses 12 technical features. 

Principal Component Analysis (PCA), Genetic Algorithms (GA), and Decision Trees (CART) are all compared in the research article \cite{tsai2010combining} by Tsai and et al. It examines their prediction accuracy and mistakes by combining them using union, intersection, and multi intersection methods. The findings of the experiments indicate that integrating several feature selection techniques may improve prediction performance over single feature selection methods. The intersection of PCA and GA, as well as the multi-intersection of PCA, GA, and CART, perform the best, with accuracy rates of 79 percent and 78.98 percent, respectively.

The causal feature selection (CFS) method is proposed in this research \cite{zhang2014causal} by Zhang and et al., to choose more representative features for improved stock prediction modeling. Comparative tests were performed between CFS and three well-known feature selection methods, namely principal component analysis (PCA), decision trees (DT; CART), and the least absolute shrinkage and selection operator, using 13-year data from the Shanghai Stock Exchanges (LASSO). When coupled with each of the seven baseline models, CFS performs best in terms of accuracy and precision in most instances and finds 18 key consistent characteristics out the 50 initial input features given.

\chapter{Research Plan}

\section{Real-time Stock Trading Strategy}

The system is built using a Quantopian working environment. It provides a large variety of financial data of major US stocks, starting from 2002 to the current date. Quantopian has many factors at its disposal for us to use and is also flexible enough to let us create our CustomFactor. These factors are a necessity for predicting the future market price using any Machine learning algorithm.

\begin{figure}[htbp]
\centering
\includegraphics[scale=0.5]{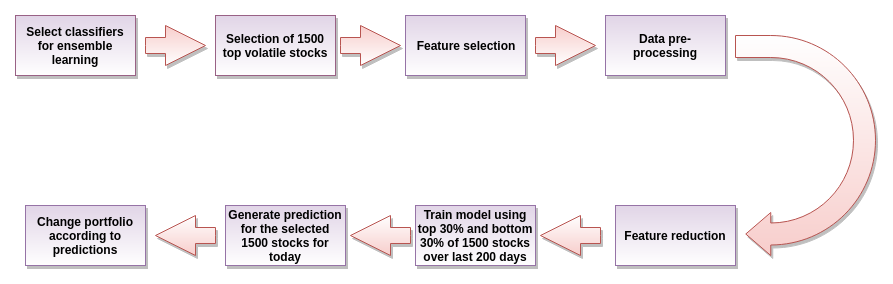}
\caption{Overall System model for Real-time trading}

\label{system}

\end{figure}

\vspace{5mm}

The diagram in the figure system shows the complete workflow of our model. The processes include selecting stocks, Feature selection, Data pre-processing, training and generating predictions using machine learning algorithms, and finally making changes to the portfolio according to the predictions.
\vspace{5mm}

Using the Quantopian algorithm feature, we implemented our strategy incorporating several machine learning algorithms into it. We started initially by collecting the data of 1500 of the top stocks in the market using the Q1500US function provided by Quantopian. Next, we imported all the factors provided whilst also including factors from TA-LIB (a significant financial factor provider ). We also had to implement some custom factors ourselves. Some of the factors are Asset Growth 3M, Asset to Equity Ratio, Capex to Cash Flows, EBIT to Assets, EBITDA Yield, Earnings Quality, MACD Signal Line, Mean Reversion 1M, AD, ADX, APO, ATR, BETA, MFI, etc. resulting in a total of 148 features.
\vspace{5mm}

But what we need to realize is that due to the volatility of the market, just feeding all the factors into the ML algorithm won’t give a very consistent result overall because a given factor can have both a positive or a negative effect on the prediction itself at the different given time concerning the market. To overcome this problem, we had to implement feature reduction dynamically, which is discussed in the \ref{featureselect} section. After the initial feature selection, we further select the top 25 features based on the F-value of ANOVA.
\vspace{5mm}

After collecting the features, we set the number of stocks we wanted to trade, Machine learning window length, Nth forward day we wanted to predict, i.e. in this case of weekly trading, we had the variable set to 5, and also the trading frequency, i.e. the number of days after which we wanted to initiate the trade. From the 1500 stock data that we imported before, we sort and only trade on two different quantiles, upper 30\% and lower 30\%. We perform this slicing to make sure that we do not trade on stocks that have a very steady rate of change on their pricing, but only trade on stocks that are placed higher and lower down the ladder on which we could go long( upper 30\%) and short( lower 30\%) and have a significant success rate on it. We set the upper 30\% to 1 i.e. long, lower 30\% to -1 i.e. short and other 40\% to 0 i.e. we do not perform any trade on them. The summation of these upper and lower quantiles results in 500 stocks, i.e. the number we set earlier. We had to strip the Label ( Returns ) from the zipline and perform a five-day computation on it. The T - 5 days data had to be discarded because there are no five days in forwarding time data for that given particular time resulting in NAN labels and thus was dropped from the zipline data frame. The Label column had to be kept separate to pass it onto the ML algorithm. Since Quantopian does not support machine learning and data preprocessing to be done inside the pipeline, we had to sort its entirety outside.
\vspace{5mm}

After preprocessing of the data, we make a new column in the pipeline called ML and call the Machine learning function to fill it up for each and every stock for that given day. The parameters of the ML function is the universe and all the columns of the pipeline, i.e. factors and label that we calculated. Here is the part where we perform the factor reduction that was talked about earlier. This process is performed dynamically throughout the training process, i.e. every single time we train the algorithm, we only train it with the top 25 features using the SelectKBest feature selection method.

\vspace{5mm}

\begin{figure}[htbp]
\centering
\includegraphics[scale=0.4]{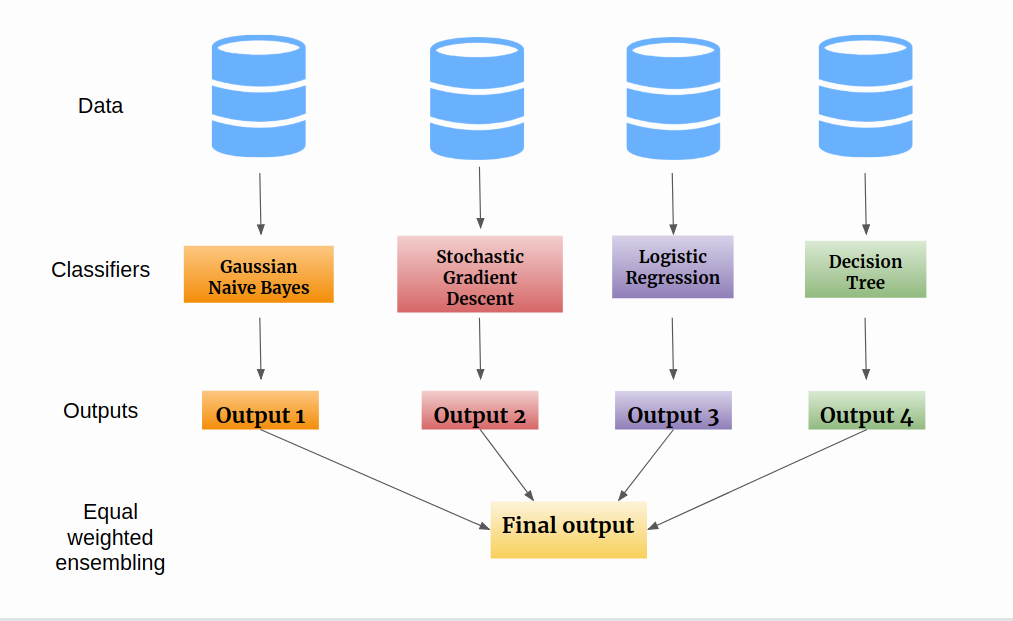}
\caption{Structure of the Machine Learning Model Used }
\label{ensemble}
\end{figure}

\vspace{5mm}

The figure \ref{ensemble} illustrates our Ensemble learning model. Here four machine learning algorithms each give us a hypothesis which each gives us an output. These four outputs are used for equal-weighted ensembling to generate the final output. An initial problem that we faced while implementing ML algo is that whenever we tried to have three or more high complexity classifiers(SVM, AdaBoost, etc.) along with the dynamic feature selection, we use to run into TLE. But then we selected algorithms that have very small runtime, usually around 2-3 seconds to test and train, which is a very important thing to have in a live trading algorithm.

\chapter{Data Processing }
\section{Proposed Feature Selection Model}
\label{featureselect}
The paper analyzes a total of 148 features based on four criteria. 

\begin{figure}[H]
\centering
\includegraphics[scale=0.65]{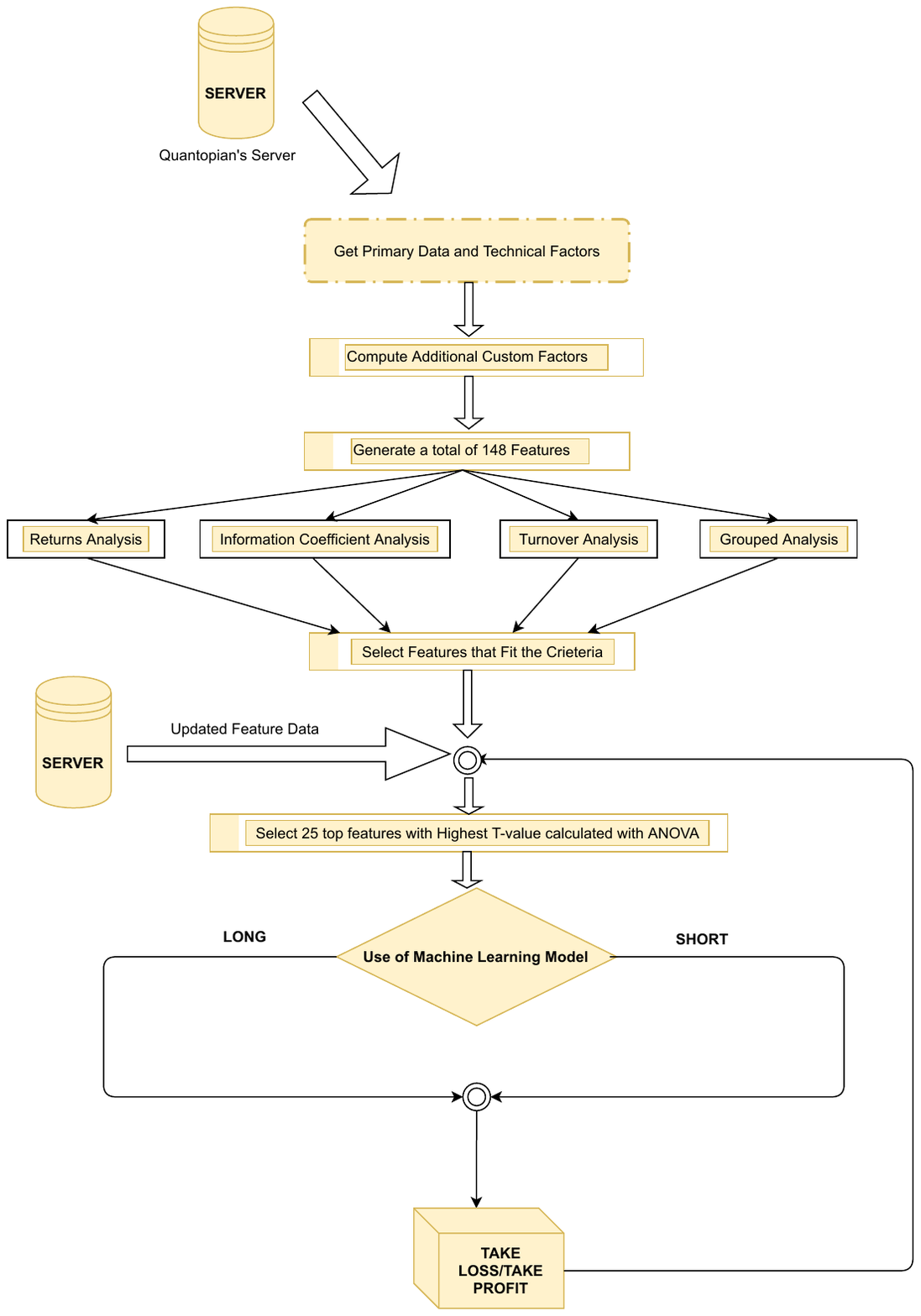}
\caption{\textcolor{blue}{Proposed Feature Selection Model}}
\label{featureSelect}
\end{figure}

\begin{itemize}
\item Returns Analysis
\item Information Coefficient Analysis
\item Turnover Analysis
\item Grouped Analysis
\end{itemize}
For the analysis we used Alphalens on the stock data from start\_date='2011-03-06'and end\_date='2012-03-06'. The feature selection method is showed in the figure\ref{featureSelect}. The method went on long and short positions on the top and bottom quantile or reverse if the feature is negative. The trading is done for 1D (1 day hold period), 5D (5 day hold period) and 22D(22 day hold period). The features must have "mean return" greater than 0.05\% or 0.5 basis points for both the long and short positions selected to trade for that certain time period. The Information coefficient must be greater than 0.005. In the turnover analysis, the mean turnover must be greater than 0.25. The stocks satisfying these criteria will be initially selected. In addition, Sklearn's SelectKbest method is used to select the best features out of the selected feature. For the hyper-parameter, "f\_classif" is selected, which ranks the features using the T-scores from ANOVA.

The stock values were divided into 3 equal quantiles. The lower quantile, the middle quantile and the upper quantile. Each quantile had about 33.33\% of the values. . Table \ref{feature1} and \ref{feature2}shows the selected features for daily, weekly and monthly trading.

\begin{table}[H]
\caption{\textcolor{blue}{Selected Features for trading based on feature analysis for different timeframe (Part 1)}}
\begin{tabular}{llllllll}

\hline
\rowcolor[HTML]{FFCE93} 
\multicolumn{1}{|l|}{\cellcolor[HTML]{FFCE93}{\color[HTML]{656565} }} & \multicolumn{3}{l|}{\cellcolor[HTML]{FFCE93}{\color[HTML]{656565} Trading Position}} & \multicolumn{1}{l|}{\cellcolor[HTML]{FFCE93}{\color[HTML]{656565} }} & \multicolumn{3}{l|}{\cellcolor[HTML]{FFCE93}{\color[HTML]{656565} Trading Position}} \\ \cline{2-4} \cline{6-8} 
\rowcolor[HTML]{FFCE93} 
\multicolumn{1}{|l|}{\multirow{-2}{*}{\cellcolor[HTML]{FFCE93}{\color[HTML]{656565} Feature Name}}} & \multicolumn{1}{l|}{\cellcolor[HTML]{FFCE93}{\color[HTML]{656565} 1D}} & \multicolumn{1}{l|}{\cellcolor[HTML]{FFCE93}{\color[HTML]{656565} 1W}} & \multicolumn{1}{l|}{\cellcolor[HTML]{FFCE93}{\color[HTML]{656565} 1M}} & \multicolumn{1}{l|}{\multirow{-2}{*}{\cellcolor[HTML]{FFCE93}{\color[HTML]{656565} Feature Name}}} & \multicolumn{1}{l|}{\cellcolor[HTML]{FFCE93}{\color[HTML]{656565} 1D}} & \multicolumn{1}{l|}{\cellcolor[HTML]{FFCE93}{\color[HTML]{656565} 1W}} & \multicolumn{1}{l|}{\cellcolor[HTML]{FFCE93}{\color[HTML]{656565} 1M}} \\ \hline
\multicolumn{1}{|l|}{Asset\_Growth\_5D} & \multicolumn{1}{l|}{} & \multicolumn{1}{l|}{} & \multicolumn{1}{l|}{} & \multicolumn{1}{l|}{BETA\_6D} & \multicolumn{1}{l|}{} & \multicolumn{1}{l|}{} & \multicolumn{1}{l|}{} \\ \hline
\multicolumn{1}{|l|}{Asset\_Growth\_2D} & \multicolumn{1}{l|}{\checkmark} & \multicolumn{1}{l|}{\checkmark} & \multicolumn{1}{l|}{\checkmark} & \multicolumn{1}{l|}{BETA\_1D} & \multicolumn{1}{l|}{} & \multicolumn{1}{l|}{} & \multicolumn{1}{l|}{} \\ \hline
\multicolumn{1}{|l|}{Asset\_Growth\_5D} & \multicolumn{1}{l|}{} & \multicolumn{1}{l|}{} & \multicolumn{1}{l|}{} & \multicolumn{1}{l|}{BETA\_2D} & \multicolumn{1}{l|}{} & \multicolumn{1}{l|}{} & \multicolumn{1}{l|}{} \\ \hline
\multicolumn{1}{|l|}{Asset\_Growth\_22D} & \multicolumn{1}{l|}{} & \multicolumn{1}{l|}{} & \multicolumn{1}{l|}{} & \multicolumn{1}{l|}{BETA\_5D} & \multicolumn{1}{l|}{} & \multicolumn{1}{l|}{} & \multicolumn{1}{l|}{} \\ \hline
\multicolumn{1}{|l|}{Asset\_To\_Equity\_Ratio} & \multicolumn{1}{l|}{\checkmark} & \multicolumn{1}{l|}{\checkmark} & \multicolumn{1}{l|}{\checkmark} & \multicolumn{1}{l|}{BETA\_22D} & \multicolumn{1}{l|}{\checkmark} & \multicolumn{1}{l|}{\checkmark} & \multicolumn{1}{l|}{\checkmark} \\ \hline
\multicolumn{1}{|l|}{Capex\_To\_Cashflows} & \multicolumn{1}{l|}{\checkmark} & \multicolumn{1}{l|}{\checkmark} & \multicolumn{1}{l|}{\checkmark} & \multicolumn{1}{l|}{RSI\_10D} & \multicolumn{1}{l|}{} & \multicolumn{1}{l|}{} & \multicolumn{1}{l|}{\checkmark} \\ \hline
\multicolumn{1}{|l|}{EBITDA\_Yield} & \multicolumn{1}{l|}{} & \multicolumn{1}{l|}{\checkmark} & \multicolumn{1}{l|}{} & \multicolumn{1}{l|}{BOP} & \multicolumn{1}{l|}{\checkmark} & \multicolumn{1}{l|}{} & \multicolumn{1}{l|}{} \\ \hline
\multicolumn{1}{|l|}{EBIT\_To\_Assets} & \multicolumn{1}{l|}{\checkmark} & \multicolumn{1}{l|}{\checkmark} & \multicolumn{1}{l|}{\checkmark} & \multicolumn{1}{l|}{CCI\_14D} & \multicolumn{1}{l|}{} & \multicolumn{1}{l|}{} & \multicolumn{1}{l|}{\checkmark} \\ \hline
\multicolumn{1}{|l|}{Net\_Income\_Margin} & \multicolumn{1}{l|}{\checkmark} & \multicolumn{1}{l|}{\checkmark} & \multicolumn{1}{l|}{\checkmark} & \multicolumn{1}{l|}{CCI\_1D} & \multicolumn{1}{l|}{} & \multicolumn{1}{l|}{} & \multicolumn{1}{l|}{} \\ \hline
\multicolumn{1}{|l|}{Return\_On\_Invest\_Capital} & \multicolumn{1}{l|}{\checkmark} & \multicolumn{1}{l|}{\checkmark} & \multicolumn{1}{l|}{\checkmark} & \multicolumn{1}{l|}{CCI\_2D} & \multicolumn{1}{l|}{\checkmark} & \multicolumn{1}{l|}{} & \multicolumn{1}{l|}{} \\ \hline
\multicolumn{1}{|l|}{Mean\_Reversion\_1M} & \multicolumn{1}{l|}{} & \multicolumn{1}{l|}{} & \multicolumn{1}{l|}{\checkmark} & \multicolumn{1}{l|}{CCI\_5D} & \multicolumn{1}{l|}{\checkmark} & \multicolumn{1}{l|}{\checkmark} & \multicolumn{1}{l|}{} \\ \hline
\multicolumn{1}{|l|}{Mean\_Reversion\_2D} & \multicolumn{1}{l|}{\checkmark} & \multicolumn{1}{l|}{} & \multicolumn{1}{l|}{} & \multicolumn{1}{l|}{CCI\_22D} & \multicolumn{1}{l|}{} & \multicolumn{1}{l|}{} & \multicolumn{1}{l|}{} \\ \hline
\multicolumn{1}{|l|}{Mean\_Reversion\_5D} & \multicolumn{1}{l|}{} & \multicolumn{1}{l|}{} & \multicolumn{1}{l|}{} & \multicolumn{1}{l|}{CMO\_15D} & \multicolumn{1}{l|}{} & \multicolumn{1}{l|}{} & \multicolumn{1}{l|}{\checkmark} \\ \hline
\multicolumn{1}{|l|}{Mean\_Reversion\_6D} & \multicolumn{1}{l|}{} & \multicolumn{1}{l|}{\checkmark} & \multicolumn{1}{l|}{} & \multicolumn{1}{l|}{CMO\_5D} & \multicolumn{1}{l|}{} & \multicolumn{1}{l|}{} & \multicolumn{1}{l|}{} \\ \hline
\multicolumn{1}{|l|}{MACD\_Signal\_10d} & \multicolumn{1}{l|}{\checkmark} & \multicolumn{1}{l|}{\checkmark} & \multicolumn{1}{l|}{\checkmark} & \multicolumn{1}{l|}{CMO\_2D} & \multicolumn{1}{l|}{} & \multicolumn{1}{l|}{} & \multicolumn{1}{l|}{} \\ \hline
\multicolumn{1}{|l|}{MACD\_Signal\_1d} & \multicolumn{1}{l|}{} & \multicolumn{1}{l|}{} & \multicolumn{1}{l|}{} & \multicolumn{1}{l|}{CMO\_22D} & \multicolumn{1}{l|}{} & \multicolumn{1}{l|}{\checkmark} & \multicolumn{1}{l|}{} \\ \hline
\multicolumn{1}{|l|}{MACD\_Signal\_2d} & \multicolumn{1}{l|}{} & \multicolumn{1}{l|}{} & \multicolumn{1}{l|}{} & \multicolumn{1}{l|}{DX\_15D} & \multicolumn{1}{l|}{} & \multicolumn{1}{l|}{} & \multicolumn{1}{l|}{} \\ \hline
\multicolumn{1}{|l|}{MACD\_Signal\_5d} & \multicolumn{1}{l|}{} & \multicolumn{1}{l|}{} & \multicolumn{1}{l|}{} & \multicolumn{1}{l|}{DX\_22D} & \multicolumn{1}{l|}{} & \multicolumn{1}{l|}{\checkmark} & \multicolumn{1}{l|}{\checkmark} \\ \hline
\multicolumn{1}{|l|}{MACD\_Signal\_22d} & \multicolumn{1}{l|}{} & \multicolumn{1}{l|}{} & \multicolumn{1}{l|}{} & \multicolumn{1}{l|}{DX\_2D} & \multicolumn{1}{l|}{\checkmark} & \multicolumn{1}{l|}{} & \multicolumn{1}{l|}{} \\ \hline
\multicolumn{1}{|l|}{AD\_14D} & \multicolumn{1}{l|}{} & \multicolumn{1}{l|}{} & \multicolumn{1}{l|}{} & \multicolumn{1}{l|}{DX\_5D} & \multicolumn{1}{l|}{} & \multicolumn{1}{l|}{} & \multicolumn{1}{l|}{} \\ \hline
\multicolumn{1}{|l|}{AD\_1D} & \multicolumn{1}{l|}{\checkmark} & \multicolumn{1}{l|}{} & \multicolumn{1}{l|}{} & \multicolumn{1}{l|}{MAX} & \multicolumn{1}{l|}{} & \multicolumn{1}{l|}{} & \multicolumn{1}{l|}{} \\ \hline
\multicolumn{1}{|l|}{AD\_2D} & \multicolumn{1}{l|}{} & \multicolumn{1}{l|}{} & \multicolumn{1}{l|}{} & \multicolumn{1}{l|}{MAXINDEX} & \multicolumn{1}{l|}{} & \multicolumn{1}{l|}{} & \multicolumn{1}{l|}{} \\ \hline
\multicolumn{1}{|l|}{AD\_5D} & \multicolumn{1}{l|}{} & \multicolumn{1}{l|}{\checkmark} & \multicolumn{1}{l|}{} & \multicolumn{1}{l|}{MEDPRICE\_1D} & \multicolumn{1}{l|}{\checkmark} & \multicolumn{1}{l|}{\checkmark} & \multicolumn{1}{l|}{\checkmark} \\ \hline
\multicolumn{1}{|l|}{AD\_22D} & \multicolumn{1}{l|}{} & \multicolumn{1}{l|}{} & \multicolumn{1}{l|}{\checkmark} & \multicolumn{1}{l|}{MEDPRICE\_2D} & \multicolumn{1}{l|}{} & \multicolumn{1}{l|}{} & \multicolumn{1}{l|}{} \\ \hline
\multicolumn{1}{|l|}{ADX\_29D} & \multicolumn{1}{l|}{\checkmark} & \multicolumn{1}{l|}{} & \multicolumn{1}{l|}{\checkmark} & \multicolumn{1}{l|}{MEDPRICE\_5D} & \multicolumn{1}{l|}{} & \multicolumn{1}{l|}{} & \multicolumn{1}{l|}{} \\ \hline
\multicolumn{1}{|l|}{ADX\_1D} & \multicolumn{1}{l|}{} & \multicolumn{1}{l|}{} & \multicolumn{1}{l|}{} & \multicolumn{1}{l|}{MEDPRICE\_22D} & \multicolumn{1}{l|}{} & \multicolumn{1}{l|}{} & \multicolumn{1}{l|}{} \\ \hline
\multicolumn{1}{|l|}{ADX\_2D} & \multicolumn{1}{l|}{} & \multicolumn{1}{l|}{} & \multicolumn{1}{l|}{} & \multicolumn{1}{l|}{MFI\_15D} & \multicolumn{1}{l|}{} & \multicolumn{1}{l|}{} & \multicolumn{1}{l|}{} \\ \hline
\multicolumn{1}{|l|}{ADX\_5D} & \multicolumn{1}{l|}{} & \multicolumn{1}{l|}{} & \multicolumn{1}{l|}{} & \multicolumn{1}{l|}{MFI\_1D} & \multicolumn{1}{l|}{} & \multicolumn{1}{l|}{} & \multicolumn{1}{l|}{} \\ \hline
\multicolumn{1}{|l|}{ADX\_22D} & \multicolumn{1}{l|}{} & \multicolumn{1}{l|}{} & \multicolumn{1}{l|}{} & \multicolumn{1}{l|}{MFI\_2D} & \multicolumn{1}{l|}{} & \multicolumn{1}{l|}{} & \multicolumn{1}{l|}{\checkmark} \\ \hline
\multicolumn{1}{|l|}{APO\_12D\_26D} & \multicolumn{1}{l|}{} & \multicolumn{1}{l|}{} & \multicolumn{1}{l|}{} & \multicolumn{1}{l|}{MFI\_5D} & \multicolumn{1}{l|}{} & \multicolumn{1}{l|}{} & \multicolumn{1}{l|}{} \\ \hline
\multicolumn{1}{|l|}{ATR\_15D} & \multicolumn{1}{l|}{} & \multicolumn{1}{l|}{\checkmark} & \multicolumn{1}{l|}{\checkmark} & \multicolumn{1}{l|}{MFI\_22D} & \multicolumn{1}{l|}{\checkmark} & \multicolumn{1}{l|}{\checkmark} & \multicolumn{1}{l|}{} \\ \hline
\multicolumn{1}{|l|}{ATR\_2D} & \multicolumn{1}{l|}{\checkmark} & \multicolumn{1}{l|}{} & \multicolumn{1}{l|}{} & \multicolumn{1}{l|}{MIDPOINT} & \multicolumn{1}{l|}{} & \multicolumn{1}{l|}{} & \multicolumn{1}{l|}{} \\ \hline
\multicolumn{1}{|l|}{ATR\_5D} & \multicolumn{1}{l|}{} & \multicolumn{1}{l|}{} & \multicolumn{1}{l|}{} & \multicolumn{1}{l|}{MIN} & \multicolumn{1}{l|}{} & \multicolumn{1}{l|}{} & \multicolumn{1}{l|}{} \\ \hline
\multicolumn{1}{|l|}{ATR\_22D} & \multicolumn{1}{l|}{} & \multicolumn{1}{l|}{} & \multicolumn{1}{l|}{} & \multicolumn{1}{l|}{MININDEX} & \multicolumn{1}{l|}{} & \multicolumn{1}{l|}{} & \multicolumn{1}{l|}{} \\ \hline
 &  &  &  &  &  &  & 
\end{tabular}

\label{feature1}
\end{table}

\begin{table}[H]
\caption{\textcolor{blue}{Selected Features for trading based on feature analysis for different timeframe (Part 2)}}
\begin{tabular}{llllllll}

\hline
\rowcolor[HTML]{FFCE93} 
\multicolumn{1}{|l|}{\cellcolor[HTML]{FFCE93}{\color[HTML]{656565} }} & \multicolumn{3}{l|}{\cellcolor[HTML]{FFCE93}{\color[HTML]{656565} Trading Position}} & \multicolumn{1}{l|}{\cellcolor[HTML]{FFCE93}{\color[HTML]{656565} }} & \multicolumn{3}{l|}{\cellcolor[HTML]{FFCE93}{\color[HTML]{656565} Trading Position}} \\ \cline{2-4} \cline{6-8} 
\rowcolor[HTML]{FFCE93} 
\multicolumn{1}{|l|}{\multirow{-2}{*}{\cellcolor[HTML]{FFCE93}{\color[HTML]{656565} Feature Name}}} & \multicolumn{1}{l|}{\cellcolor[HTML]{FFCE93}{\color[HTML]{656565} 1D}} & \multicolumn{1}{l|}{\cellcolor[HTML]{FFCE93}{\color[HTML]{656565} 1W}} & \multicolumn{1}{l|}{\cellcolor[HTML]{FFCE93}{\color[HTML]{656565} 1M}} & \multicolumn{1}{l|}{\multirow{-2}{*}{\cellcolor[HTML]{FFCE93}{\color[HTML]{656565} Feature Name}}} & \multicolumn{1}{l|}{\cellcolor[HTML]{FFCE93}{\color[HTML]{656565} 1D}} & \multicolumn{1}{l|}{\cellcolor[HTML]{FFCE93}{\color[HTML]{656565} 1W}} & \multicolumn{1}{l|}{\cellcolor[HTML]{FFCE93}{\color[HTML]{656565} 1M}} \\ \hline
\multicolumn{1}{|l|}{MINUS\_DI\_15D} & \multicolumn{1}{l|}{} & \multicolumn{1}{l|}{} & \multicolumn{1}{l|}{\checkmark} & \multicolumn{1}{l|}{WILLR\_14D} & \multicolumn{1}{l|}{} & \multicolumn{1}{l|}{} & \multicolumn{1}{l|}{\checkmark} \\ \hline
\multicolumn{1}{|l|}{MINUS\_DI\_1D} & \multicolumn{1}{l|}{} & \multicolumn{1}{l|}{} & \multicolumn{1}{l|}{} & \multicolumn{1}{l|}{WILLR\_1D} & \multicolumn{1}{l|}{\checkmark} & \multicolumn{1}{l|}{} & \multicolumn{1}{l|}{} \\ \hline
\multicolumn{1}{|l|}{MINUS\_DI\_2D} & \multicolumn{1}{l|}{} & \multicolumn{1}{l|}{} & \multicolumn{1}{l|}{} & \multicolumn{1}{l|}{WILLR\_2D} & \multicolumn{1}{l|}{} & \multicolumn{1}{l|}{} & \multicolumn{1}{l|}{} \\ \hline
\multicolumn{1}{|l|}{MINUS\_DI\_5D} & \multicolumn{1}{l|}{} & \multicolumn{1}{l|}{} & \multicolumn{1}{l|}{} & \multicolumn{1}{l|}{WILLR\_5D} & \multicolumn{1}{l|}{} & \multicolumn{1}{l|}{} & \multicolumn{1}{l|}{} \\ \hline
\multicolumn{1}{|l|}{MINUS\_DI\_22D} & \multicolumn{1}{l|}{} & \multicolumn{1}{l|}{} & \multicolumn{1}{l|}{} & \multicolumn{1}{l|}{WILLR\_22D} & \multicolumn{1}{l|}{} & \multicolumn{1}{l|}{\checkmark} & \multicolumn{1}{l|}{} \\ \hline
\multicolumn{1}{|l|}{MINUS\_DM\_15D} & \multicolumn{1}{l|}{} & \multicolumn{1}{l|}{} & \multicolumn{1}{l|}{} & \multicolumn{1}{l|}{Average\_Dollar\_Volume} & \multicolumn{1}{l|}{\checkmark} & \multicolumn{1}{l|}{\checkmark} & \multicolumn{1}{l|}{\checkmark} \\ \hline
\multicolumn{1}{|l|}{MINUS\_DM\_1D} & \multicolumn{1}{l|}{} & \multicolumn{1}{l|}{} & \multicolumn{1}{l|}{} & \multicolumn{1}{l|}{Moneyflow\_Volume\_5D} & \multicolumn{1}{l|}{} & \multicolumn{1}{l|}{} & \multicolumn{1}{l|}{} \\ \hline
\multicolumn{1}{|l|}{MINUS\_DM\_2D} & \multicolumn{1}{l|}{\checkmark} & \multicolumn{1}{l|}{} & \multicolumn{1}{l|}{} & \multicolumn{1}{l|}{Moneyflow\_Volume\_1D} & \multicolumn{1}{l|}{} & \multicolumn{1}{l|}{} & \multicolumn{1}{l|}{\checkmark} \\ \hline
\multicolumn{1}{|l|}{MINUS\_DM\_5D} & \multicolumn{1}{l|}{} & \multicolumn{1}{l|}{} & \multicolumn{1}{l|}{} & \multicolumn{1}{l|}{Moneyflow\_Volume\_2D} & \multicolumn{1}{l|}{} & \multicolumn{1}{l|}{\checkmark} & \multicolumn{1}{l|}{} \\ \hline
\multicolumn{1}{|l|}{MINUS\_DM\_22D} & \multicolumn{1}{l|}{} & \multicolumn{1}{l|}{} & \multicolumn{1}{l|}{} & \multicolumn{1}{l|}{Moneyflow\_Volume\_22D} & \multicolumn{1}{l|}{\checkmark} & \multicolumn{1}{l|}{} & \multicolumn{1}{l|}{} \\ \hline
\multicolumn{1}{|l|}{PLUS\_DI\_15D} & \multicolumn{1}{l|}{} & \multicolumn{1}{l|}{} & \multicolumn{1}{l|}{} & \multicolumn{1}{l|}{Annualized\_Volatility} & \multicolumn{1}{l|}{\checkmark} & \multicolumn{1}{l|}{\checkmark} & \multicolumn{1}{l|}{\checkmark} \\ \hline
\multicolumn{1}{|l|}{PLUS\_DI\_1D} & \multicolumn{1}{l|}{} & \multicolumn{1}{l|}{} & \multicolumn{1}{l|}{} & \multicolumn{1}{l|}{Operating\_Cashflows\_To\_Assets} & \multicolumn{1}{l|}{\checkmark} & \multicolumn{1}{l|}{\checkmark} & \multicolumn{1}{l|}{\checkmark} \\ \hline
\multicolumn{1}{|l|}{PLUS\_DI\_2D} & \multicolumn{1}{l|}{\checkmark} & \multicolumn{1}{l|}{} & \multicolumn{1}{l|}{} & \multicolumn{1}{l|}{Price\_Momentum\_3M} & \multicolumn{1}{l|}{\checkmark} & \multicolumn{1}{l|}{} & \multicolumn{1}{l|}{\checkmark} \\ \hline
\multicolumn{1}{|l|}{PLUS\_DI\_5D} & \multicolumn{1}{l|}{} & \multicolumn{1}{l|}{\checkmark} & \multicolumn{1}{l|}{\checkmark} & \multicolumn{1}{l|}{Price\_Oscillator\_20D} & \multicolumn{1}{l|}{} & \multicolumn{1}{l|}{} & \multicolumn{1}{l|}{} \\ \hline
\multicolumn{1}{|l|}{PLUS\_DI\_22D} & \multicolumn{1}{l|}{} & \multicolumn{1}{l|}{\checkmark} & \multicolumn{1}{l|}{} & \multicolumn{1}{l|}{Price\_Oscillator\_1D} & \multicolumn{1}{l|}{\checkmark} & \multicolumn{1}{l|}{} & \multicolumn{1}{l|}{} \\ \hline
\multicolumn{1}{|l|}{PLUS\_DM\_15D} & \multicolumn{1}{l|}{} & \multicolumn{1}{l|}{} & \multicolumn{1}{l|}{} & \multicolumn{1}{l|}{Price\_Oscillator\_2D} & \multicolumn{1}{l|}{} & \multicolumn{1}{l|}{} & \multicolumn{1}{l|}{} \\ \hline
\multicolumn{1}{|l|}{PLUS\_DM\_1D} & \multicolumn{1}{l|}{} & \multicolumn{1}{l|}{} & \multicolumn{1}{l|}{\checkmark} & \multicolumn{1}{l|}{Price\_Oscillator\_5D} & \multicolumn{1}{l|}{} & \multicolumn{1}{l|}{} & \multicolumn{1}{l|}{} \\ \hline
\multicolumn{1}{|l|}{PLUS\_DM\_2D} & \multicolumn{1}{l|}{} & \multicolumn{1}{l|}{} & \multicolumn{1}{l|}{} & \multicolumn{1}{l|}{Price\_Oscillator\_22D} & \multicolumn{1}{l|}{} & \multicolumn{1}{l|}{} & \multicolumn{1}{l|}{} \\ \hline
\multicolumn{1}{|l|}{PLUS\_DM\_5D} & \multicolumn{1}{l|}{} & \multicolumn{1}{l|}{} & \multicolumn{1}{l|}{} & \multicolumn{1}{l|}{Returns\_215D} & \multicolumn{1}{l|}{} & \multicolumn{1}{l|}{\checkmark} & \multicolumn{1}{l|}{\checkmark} \\ \hline
\multicolumn{1}{|l|}{PLUS\_DM\_22D} & \multicolumn{1}{l|}{} & \multicolumn{1}{l|}{} & \multicolumn{1}{l|}{} & \multicolumn{1}{l|}{Returns\_190D} & \multicolumn{1}{l|}{} & \multicolumn{1}{l|}{} & \multicolumn{1}{l|}{} \\ \hline
\multicolumn{1}{|l|}{PPO\_12D\_26D} & \multicolumn{1}{l|}{\checkmark} & \multicolumn{1}{l|}{} & \multicolumn{1}{l|}{} & \multicolumn{1}{l|}{Returns\_160D} & \multicolumn{1}{l|}{} & \multicolumn{1}{l|}{} & \multicolumn{1}{l|}{} \\ \hline
\multicolumn{1}{|l|}{PPO\_8D\_13D} & \multicolumn{1}{l|}{} & \multicolumn{1}{l|}{\checkmark} & \multicolumn{1}{l|}{} & \multicolumn{1}{l|}{Returns\_100D} & \multicolumn{1}{l|}{\checkmark} & \multicolumn{1}{l|}{} & \multicolumn{1}{l|}{} \\ \hline
\multicolumn{1}{|l|}{PPO\_1D\_3D} & \multicolumn{1}{l|}{} & \multicolumn{1}{l|}{} & \multicolumn{1}{l|}{} & \multicolumn{1}{l|}{Returns\_50D} & \multicolumn{1}{l|}{} & \multicolumn{1}{l|}{} & \multicolumn{1}{l|}{} \\ \hline
\multicolumn{1}{|l|}{PPO\_24D\_50D} & \multicolumn{1}{l|}{} & \multicolumn{1}{l|}{} & \multicolumn{1}{l|}{\checkmark} & \multicolumn{1}{l|}{Returns\_25D} & \multicolumn{1}{l|}{} & \multicolumn{1}{l|}{} & \multicolumn{1}{l|}{} \\ \hline
\multicolumn{1}{|l|}{STDDEV} & \multicolumn{1}{l|}{} & \multicolumn{1}{l|}{} & \multicolumn{1}{l|}{} & \multicolumn{1}{l|}{Trendline\_252D} & \multicolumn{1}{l|}{\checkmark} & \multicolumn{1}{l|}{\checkmark} & \multicolumn{1}{l|}{\checkmark} \\ \hline
\multicolumn{1}{|l|}{TRANGE\_2D} & \multicolumn{1}{l|}{} & \multicolumn{1}{l|}{\checkmark} & \multicolumn{1}{l|}{\checkmark} & \multicolumn{1}{l|}{Trendline\_25D} & \multicolumn{1}{l|}{} & \multicolumn{1}{l|}{} & \multicolumn{1}{l|}{} \\ \hline
\multicolumn{1}{|l|}{TRANGE\_1D} & \multicolumn{1}{l|}{} & \multicolumn{1}{l|}{} & \multicolumn{1}{l|}{} & \multicolumn{1}{l|}{Trendline\_50D} & \multicolumn{1}{l|}{} & \multicolumn{1}{l|}{} & \multicolumn{1}{l|}{} \\ \hline
\multicolumn{1}{|l|}{TRANGE\_5D} & \multicolumn{1}{l|}{\checkmark} & \multicolumn{1}{l|}{} & \multicolumn{1}{l|}{} & \multicolumn{1}{l|}{Trendline\_100D} & \multicolumn{1}{l|}{} & \multicolumn{1}{l|}{} & \multicolumn{1}{l|}{} \\ \hline
\multicolumn{1}{|l|}{TRANGE\_22D} & \multicolumn{1}{l|}{} & \multicolumn{1}{l|}{} & \multicolumn{1}{l|}{} & \multicolumn{1}{l|}{Trendline\_150D} & \multicolumn{1}{l|}{} & \multicolumn{1}{l|}{} & \multicolumn{1}{l|}{} \\ \hline
\multicolumn{1}{|l|}{TYPPRICE\_1D} & \multicolumn{1}{l|}{} & \multicolumn{1}{l|}{} & \multicolumn{1}{l|}{} & \multicolumn{1}{l|}{Vol\_3M} & \multicolumn{1}{l|}{} & \multicolumn{1}{l|}{} & \multicolumn{1}{l|}{} \\ \hline
\multicolumn{1}{|l|}{TYPPRICE\_2D} & \multicolumn{1}{l|}{} & \multicolumn{1}{l|}{} & \multicolumn{1}{l|}{} & \multicolumn{1}{l|}{Vol\_1D} & \multicolumn{1}{l|}{} & \multicolumn{1}{l|}{} & \multicolumn{1}{l|}{} \\ \hline
\multicolumn{1}{|l|}{TYPPRICE\_5D} & \multicolumn{1}{l|}{} & \multicolumn{1}{l|}{} & \multicolumn{1}{l|}{} & \multicolumn{1}{l|}{Vol\_2D} & \multicolumn{1}{l|}{} & \multicolumn{1}{l|}{} & \multicolumn{1}{l|}{} \\ \hline
\multicolumn{1}{|l|}{TYPPRICE\_22D} & \multicolumn{1}{l|}{} & \multicolumn{1}{l|}{} & \multicolumn{1}{l|}{} & \multicolumn{1}{l|}{Vol\_5D} & \multicolumn{1}{l|}{\checkmark} & \multicolumn{1}{l|}{} & \multicolumn{1}{l|}{} \\ \hline
\multicolumn{1}{|l|}{Earnings\_Quality} & \multicolumn{1}{l|}{} & \multicolumn{1}{l|}{} & \multicolumn{1}{l|}{} & \multicolumn{1}{l|}{Vol\_22D} & \multicolumn{1}{l|}{} & \multicolumn{1}{l|}{\checkmark} & \multicolumn{1}{l|}{\checkmark} \\ \hline
 &  &  &  &  &  &  & 

\end{tabular}

\label{feature2}
\end{table}

\subsection{Feature Evaluation example for WILLR 14 Day}
The same evaluation was done for all the features. For demonstrating purposes we only show the graphical results of the feature WILLR\_14D. 

\begin{enumerate}
\item \textbf{Mean Period Wise Return By Factor Quartile}

\begin{figure}[H]
\centering
\includegraphics[scale=0.7]{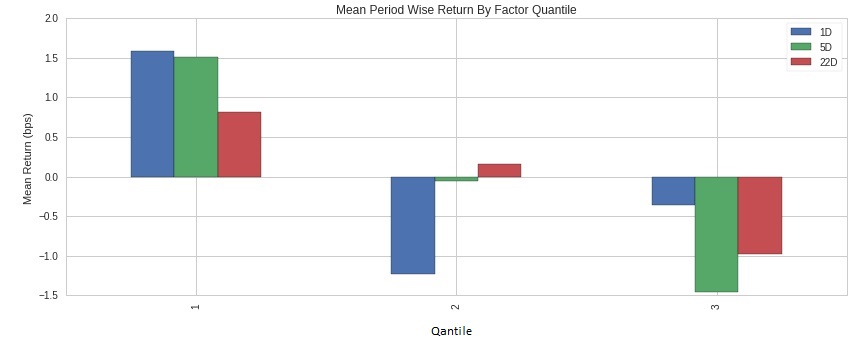}
\caption{Mean Period Wise Return By Factor Quartile}
\label{p3}
\end{figure}

The diagram in figure~\ref{p3} represents the total return as per graph height. A positive graph represents long and negative short. In Our case, we are taking 3 quartiles and breaking them into 3 separate days 1D 5D 22D, for which we can trade using the Quantopian environment. As per figure~\ref{p3} what we can use this to observe for the factor taken works best with 5D trading as we have a good amount of return for both long and short, whereas for 1d trading we can see that for the first quartile long gives a good result but is not the best for going short as displayed in the third quartile.

\item \textbf{Factor Weighted Long-Short Portfolio Cumulative Return:}

This graph represents the position of the portfolio of the trader given that person only traded taking that experimented factor into consideration alone. This represents the cumulative Returns on the portfolio of the trader.

The graphs in figure~\ref{p4} display different positions of the portfolio given 3 different trading frequencies 1D, 5D, 22D as per quartile deceleration.

\begin{figure}[H]
\centering
\includegraphics[scale=0.65]{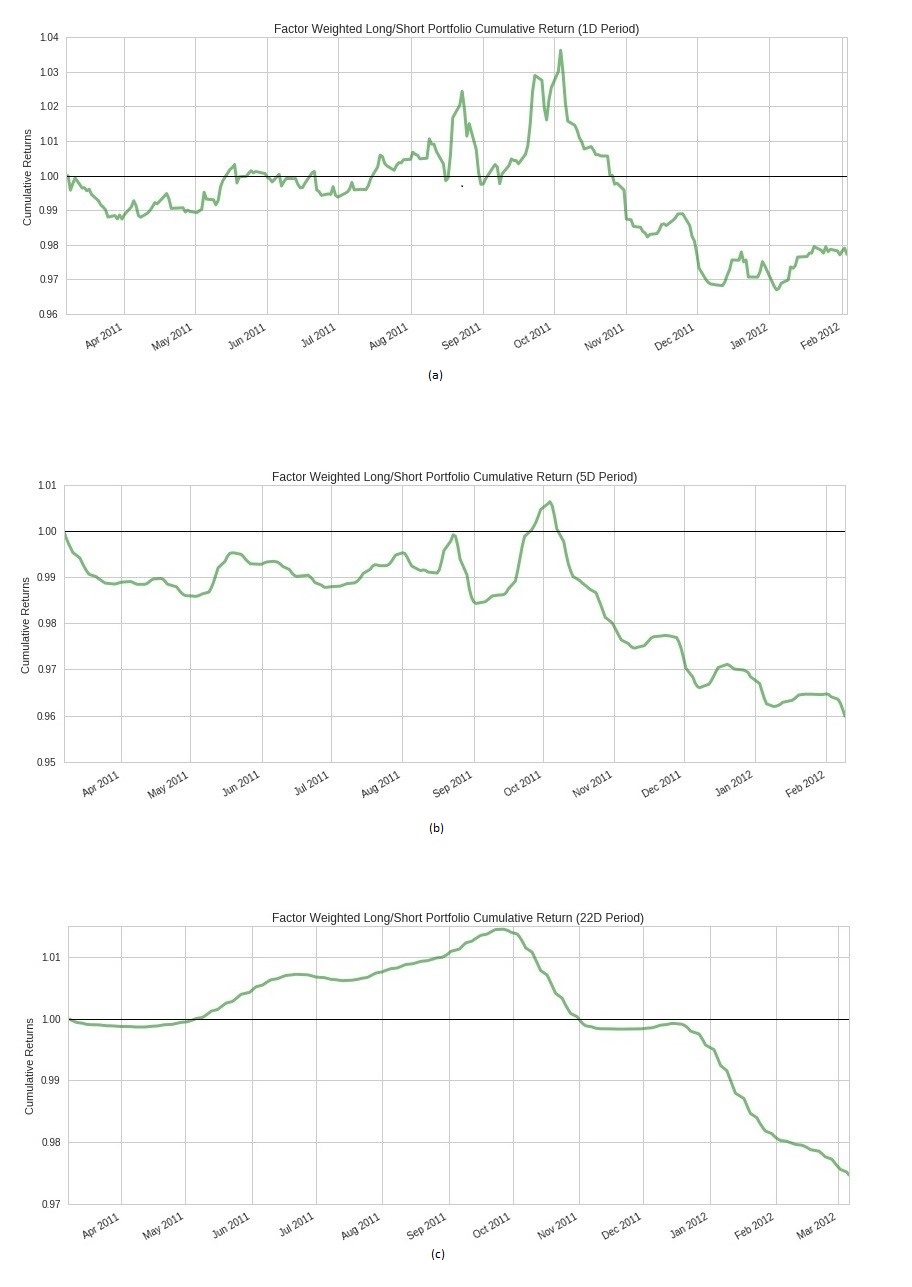}
\caption{Factor Weighted Long-Short Portfolio Cumulative Return (a)1D, (b)5D, (c)22D}
\label{p4}
\end{figure}

\item \textbf{Period Wise Return By Factor Quantile}

This graph is famously known as violin graph and comes well in handy when only the median value is not a reliable option to use in order to judge the state of the data being experimented on.

This graph is very convenient when it comes to comparing the summary statistics of the range of quartiles. The representation of this graph is very similar to that shown in the figure~\ref{p5}, but this time we get an idea of the density of where our returns are concentrated for each time period.

\begin{figure}[H]
\centering
\includegraphics[width=17cm, height=15cm]{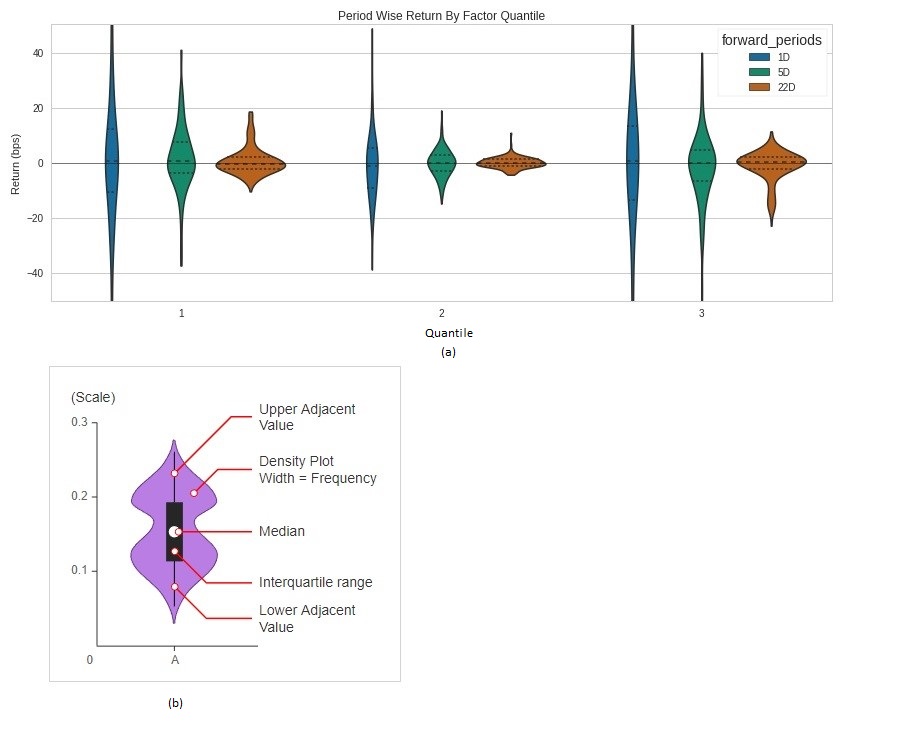}
\caption{Period Wise Return By Factor Quantile}
\label{p5}
\end{figure}

\item \textbf{Cumulative Returns by Quantile}

The cumulative quantiles of each time period are taken and are aggregated over the period of trading time. The main objective of this curve is to see of the quartiles spreads as far away from each other as possible. The far apart they are the better. The third quantile is very clearly above the first quartile and this gets more and more clearer as we move forward into time.
The less overlapping between the graphs the better.

This is calculated for the 3 different quartiles over the period of time that we traded shown in figure~\ref{p6}.

\begin{figure}[H]
\centering
\includegraphics[scale=0.7]{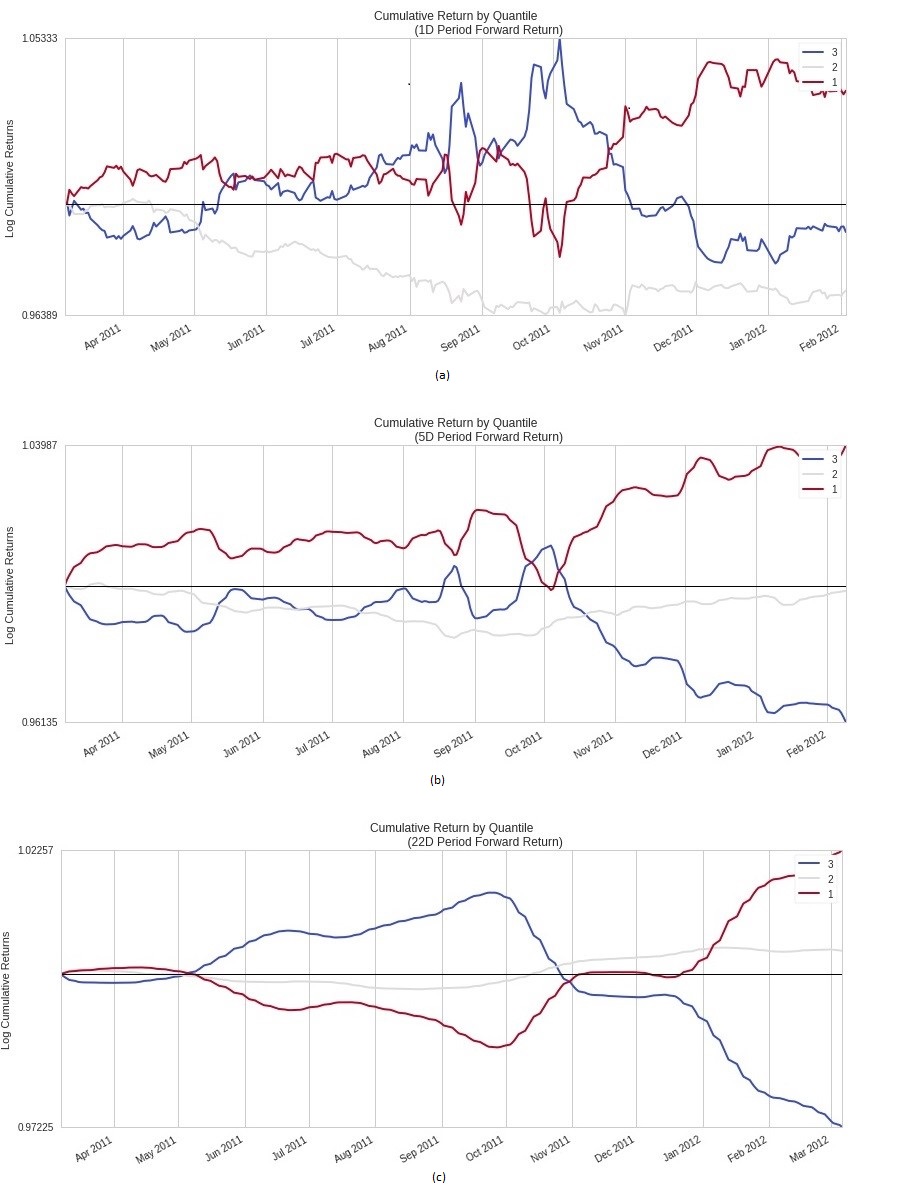}
\caption{Cumulative Returns by Quantile (a)1D, (b)5D, (c)22D}
\label{p6}
\end{figure}

\item \textbf{Top Minus Bottom Quantile Mean}


This graph in figure~\ref{p7} subtracts the top quantile from the bottom quantile and takes a mean of the answer to smoothen out the results for the given trading time period. The more positive the graph plot the more return we get over that period of trade time.

\begin{figure}[H]
\centering
\includegraphics[scale=0.65]{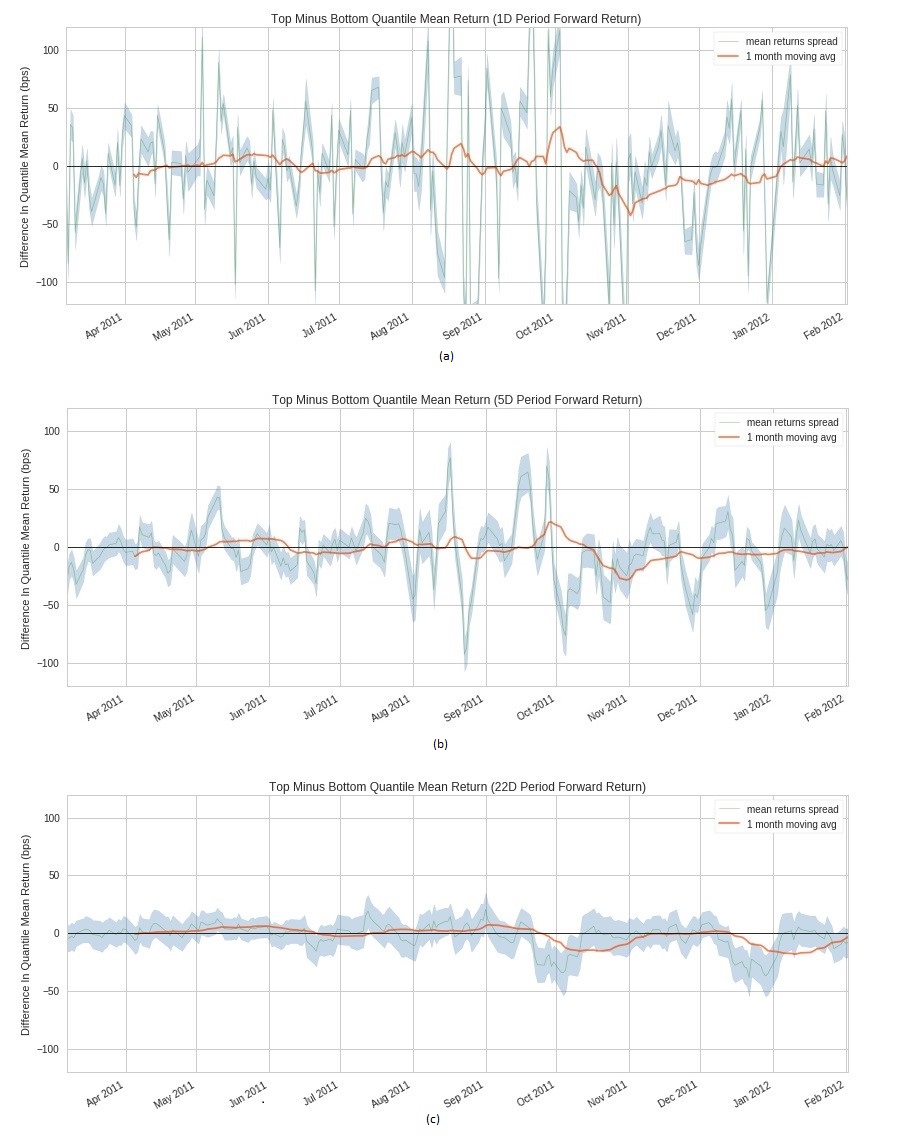}
\caption{Top Minus Bottom Quantile Mean 1D, 5D, 22D}
\label{p7}
\end{figure}

\end{enumerate}


\chapter{Results Analysis}
\section{Live Trading Results}
We used all the seven classifiers discussed to start to perform calculations on data from 2011-03-06 to 2011-09-7. We split by 80:20 ratio to form the train set and test set. Table~\ref{t3}  shows that ensemble methods work far better in this case. However, for ensemble methods, we only predicted the top and bottom values, as in real life we do not need to trade all the 1500 stocks. The ensemble 1 model is showing an accuracy of 99.25\% included LR, Gaussian\_NB, Bernoulli\_NB and SGDC whereas the ensemble 2 showing an accuracy of 74.23\% consisted of LR\_L1Regress, LR\_L2Regress, Gaussian\_NB and Bernoulli\_NB. 

\begin{table}[H]
\caption{Accuracy test on data from 1500 US stocks 2011-03-06 to 2011-09-7}
\begin{tabular}{ll}
\hline
Name of the Algorithm                                                          & Test accuracy \\ \hline
Naive Bayes(NB)                                                                & 51.21 \%      \\
Logistic Regression(LR)                                                        & 51.77 \%      \\
\begin{tabular}[c]{@{}l@{}}Stochastic Gradient \\ Descent (SGDC)\end{tabular}  & 50.56 \%      \\ \hline
\begin{tabular}[c]{@{}l@{}}Support Vector Machine\\ (SVM)\end{tabular}         & 54.06 \%      \\
Adaboost                                                                       & 53.29 \%      \\
Random Forest                                                                  & 52.43 \%      \\ \hline
\begin{tabular}[c]{@{}l@{}}Ensemble 1 \\ (predict top and bottom)\end{tabular} & 99.25 \%      \\
\begin{tabular}[c]{@{}l@{}}Ensemble 2\\ (predict top and bottom)\end{tabular}  & 74.23 \%     
\end{tabular}

\label{t3}
\end{table}

\subsection{Day Trading}
\begin{figure}[htbp]
\centering
\includegraphics[scale=0.4]{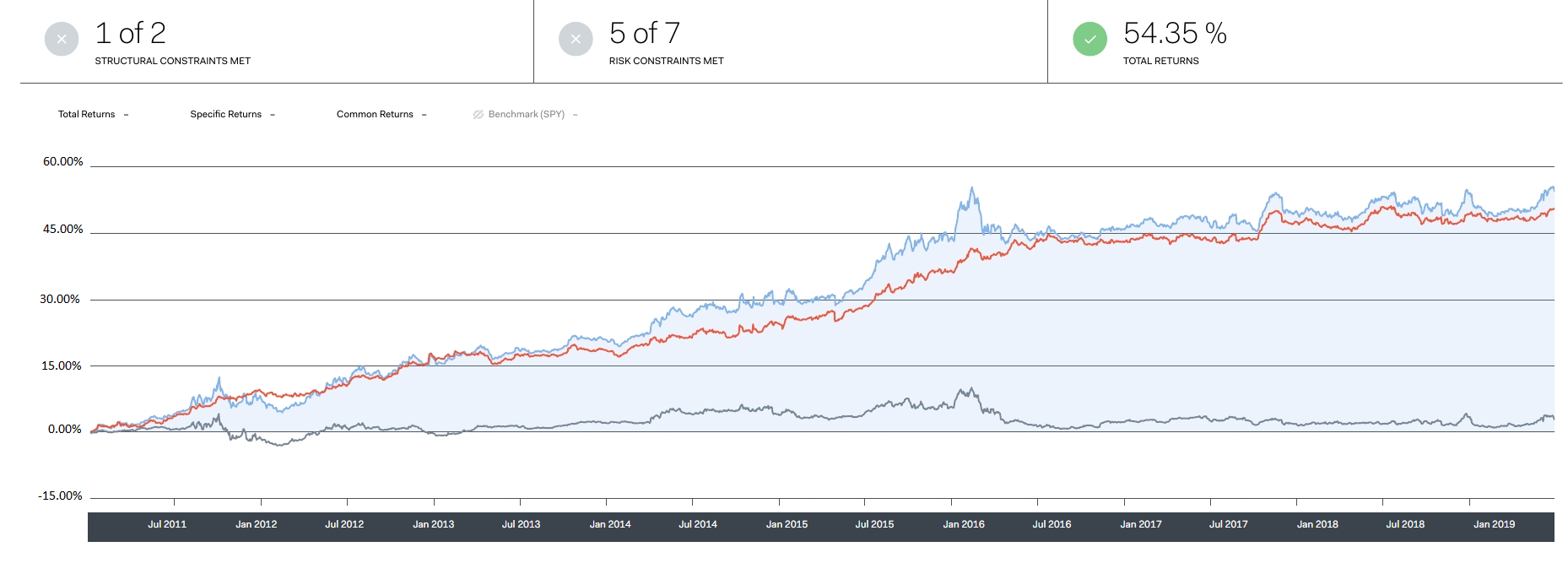}
\caption{top result Daily Trade by ensembling   GaussianNB, LogisticRegression, DTC and SGDC (best classifier)}
\label{p8}
\end{figure}

\begin{itemize}

\item \textbf{RandomForest:}
Using Random forest algorithm and daily trading we get a return of 18.08\% with a Sharpe ratio of  0.77.

\item \textbf{AdaBoost:} Using AdaBoostClassifier in the mix we get a return of 11.69\% with a sharpe ratio of 0.49.\\
\item \textbf{Ensemble 1 Classifiers:}
\begin{enumerate}
  \item GaussianNB
  \item LogisticRegression
  \item BernoulliNB
  \item Sgdc
\end{enumerate}
Using the mixed classifiers of all these algorithms together we get a return of 34.99\% with a Sharpe ratio of 0.67.
Time complexity of all these algorithms combined is very less and thus is very feasible for our purpose.

\item \textbf{Best Classifiers:}
\begin{enumerate}
  \item GaussianNB
  \item LogisticRegression
  \item DTC
  \item Sgdc
\end{enumerate}

Figure~\ref{p8} shows, using Decision tree classifiers in the mix we get a return of 54.63\% with a sharpe ratio of 1.16\%.\\

\begin{enumerate}
  \item GaussianNB
  \item LogisticRegression
  \item AdaBoostClassifier
  \item Sgdc
\end{enumerate}

\end{itemize}

\subsection{Weekly Trading:}

\begin{itemize}
    \item \textbf{AdaBoostClassifier:}
    Using AdaBoost for weekly trading we get a return of 5.25\%.

    \item \textbf{Decision Tree:}Decision tree for weekly trading we get a total return of 10.23\%.

    \item \textbf{Random Forest:}Using random forest we get a total return of 7.86\%.

\end{itemize}


\subsection{Monthly Trading:}
\begin{itemize}

    \item \textbf{AdaBoostClassifier :}
    Using AdaBoost for weekly trading we get a return of 6.16\%.

    \item \textbf{SVM :}
    Using AdaBoost for weekly trading we get a return of 13.05\%.

    \item \textbf{Random Forest :}
    Using AdaBoost for weekly trading we get a return of 4.05\%.

\end{itemize}


\subsection{Performance of our best classifier}

\textbf{Total Returns:}
It is the total amount of returns of an investment over a given period of time.This accounts for two different categories of investment.\\
\begin{enumerate}
    \item Fixed income investment
    \item Distribution and capital appreciation
\end{enumerate}

\textbf{Common Returns:} Common returns are how much of your total returns can be attributed to the common risk factors as modeled by Quantopian exposure to market beta, sectors, momentum, mean reversion, volatility, size, and value. If all your returns are common returns, it means your algorithm isn't doing anything unique and is therefore of little value. Table \ref{t2} shows 2.71\% of common returns.

\textbf{Specific Returns:}
It is an excess return that we get from an asset that is independent of specific returns of other assets. Table \ref{t2} shows 50.60\% of common returns.

\textbf{Sharpe Ratio:}
It is the measure of performance measure of investment by risk adjustment. It measures the excess returns for every unit deviation of a trade. Our approach had a 1.16\% Sharpe ratio which is decent shown in table\ref{t2}.

\begin{align}
\begin{array}{l}{\text { Sharpe Ratio }=\frac{E_{p}-E_{f}}{\sigma_{p}}} \\ {\text { where: }} \\ {E_{p}=\text { return of portfolio }} \\ {E_{f}=\text { risk-free rate }} \\ {\sigma_{p}=\text {  portfolio additional return's standard deviation}}\end{array}
\end{align}

\textbf{Max Drawdown:}
It is the maximum observed loss from the maximum observed point of the graph to the minimum point. This is used to assess the relative risk of a stock strategy.

\begin{align}
M D D=\frac{\text { Trough } \text {Value}-\text {Peak Value}}{\text {Peak Value}}
\end{align}

\textbf{Volatility:}
It is the measure of risk

\section{Performance Evaluation and Risk Evaluation:}
Our best algorithm from all of the above was the ensemble learning algorithm which incorporated Gaussian Naive Bayes classifier, Logistic Regression, decision tree classifier and Stochastic gradient descent classifier. The training day for each decision was set to be  200 days prior to that day and trading was done daily. Below are the few results that are got by running the algorithm from the date 01/04/2011 to 07/05/2019 with a capital of 10000000 USD.

\begin{table}[H]
\begin{center}
\caption{Performance of the System's Best Model  }
\begin{tabular}{ |c|c|c| } 
 \hline
 Total Returns & 54.35\% \\
 Specific Returns & 50.60\% \\
 Common Returns &2.71\% \\
 Sharp & 1.16\% \\
 Max Draw Down & -8.31\% \\
 Volatility &0.05\% \\
 \hline
\end{tabular}

\label{t2}
\end{center}
\end{table}
The table~\ref{t2} depicts that returns calculated from the initial investment were 54.35\% on the total capital. 
The average Sharpe ratio is 1.16 and the average volatility is 0.05 and the final max drawdown was -8.31. These values indicate that our model returns a portfolio that has a low level of risk.

\begin{figure}[H]
\centering
\includegraphics[scale=0.5]{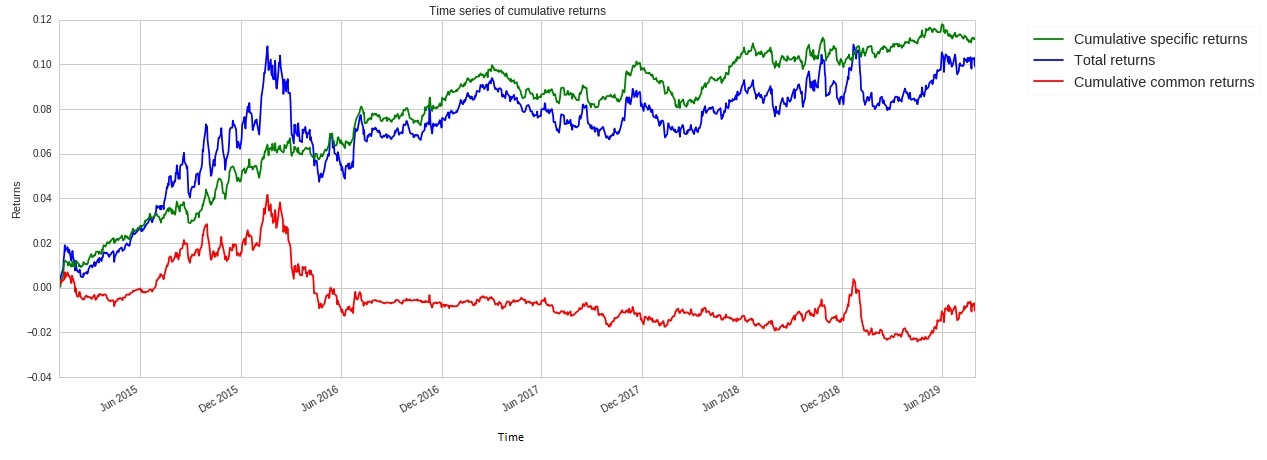}
\caption{Cumulative specific and total returns}
\label{p9}
\end{figure}

\textbf{Cumulative specific and total returns:}
Cumulative returns are independent of the time period and us the total amount of profit or loss from a particular investment. The common returns are very low which is a good sign for the model as it means that our algorithm has a low beta and performs well irrespective of whether the stock prices rise or fall. Which made the specific return very high (50.60\%) as shown in figure~\ref{p9}.

\begin{figure}[H]
\centering
\includegraphics[scale=0.7]{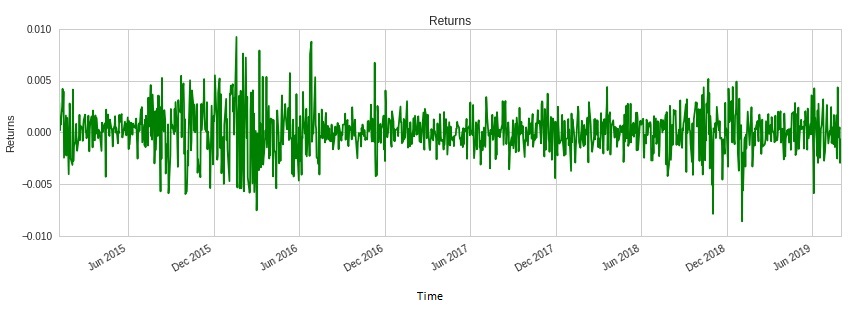}
\caption{Returns over time}
\label{p10}
\end{figure}

\textbf{Returns over time : }
Returns are gains or losses made by a particular investment. Returns can be expressed as the percentage increase or decrease in a particular investment or it can be quantified in a particular currency. The figure~\ref{p10} shows that the returns are mostly positive.

\begin{figure}[H]
\centering
\includegraphics[scale=0.7]{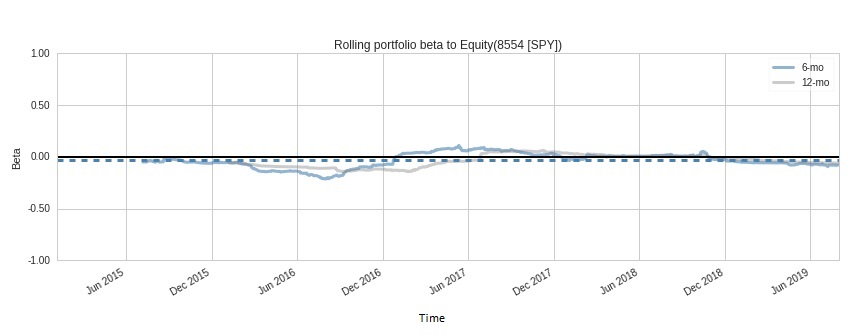}
\caption{Rolling portfolio beta to Equity}
\label{p11}
\end{figure}

\textbf{Rolling portfolio beta to Equity:}
This is shown in figure~\ref{p11}. The beta is the risk that can be attributed to the movement of the market. A beta having the value 1 signifies that a portfolio follows the trend of the market precisely. Whereas, a beta having a lower value than 1 means that a portfolio is less correlated with the overall market. A low beta value incorporated with a high Alpha value will mean that the portfolio will make a profit irrespective of the market movement.

\begin{figure}[H]
\centering
\includegraphics[scale=0.7]{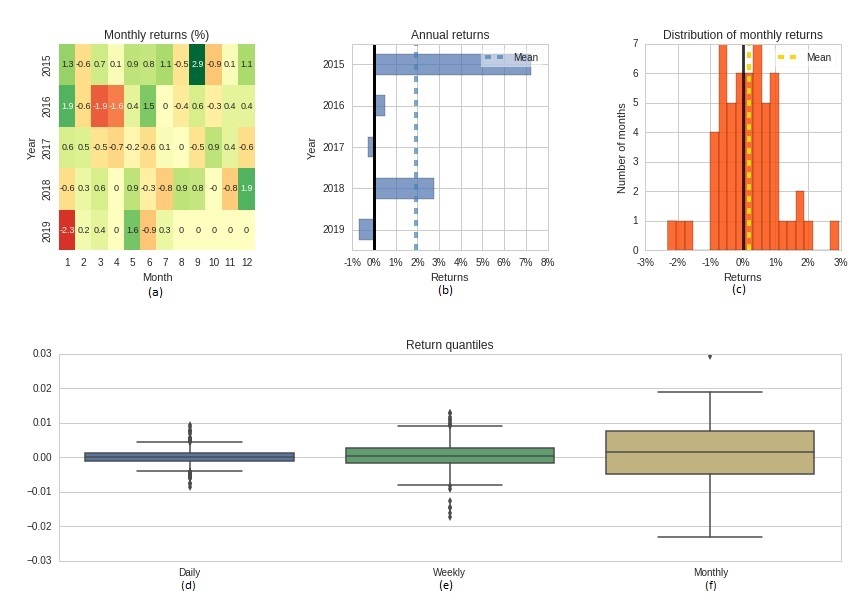}
\caption{(a)Daily returns, (b) weekly returns, (c)Monthly returns, (d)daily, (e)weekly and (f) monthly quantiles}

\label{p12}
\end{figure}

\textbf{Daily weekly and Monthly returns :}
Figure~\ref{p12} illustrates returns over the daily, weekly and monthly periods are indicated in the above figure. Each figure gives how much profit was made in a particular period.
The daily, annual  and monthly 

\textbf{Style Exposure :}
Figure~\ref{p13} shows, exposure to various investing styles. The values displayed are the rolling 63-day mean. The relevant styles are described below:

\begin{figure}[H]
\centering
\includegraphics[scale=0.8]{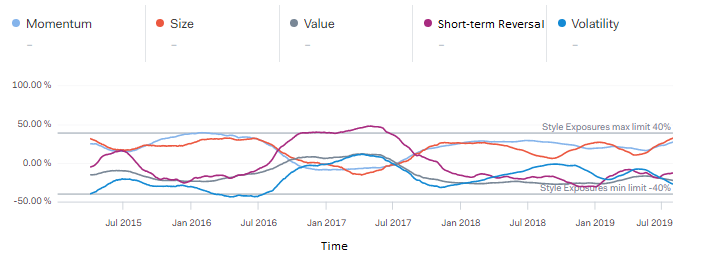}
\caption{Expose to Momentum, Size, Value, Short-term Reversal and Volatility}

\label{p13}
\end{figure}

\begin{figure}[H]
\centering
\includegraphics[scale=0.7]{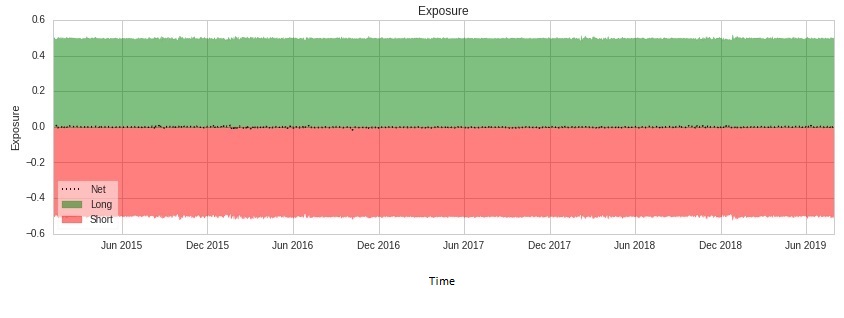}
\caption{Ratio of long and short position}

\label{p14}
\end{figure}

\textbf{Ratio of long and short position : }
We implemented an equal amount of long and short position strategy as shown in figure~\ref{p14}. So at a time, we went long on 250 stocks and short on 250 stocks. This made our model perform well both on a bull market and a bear market.

\begin{figure}[H]
\centering
\includegraphics[scale=0.7]{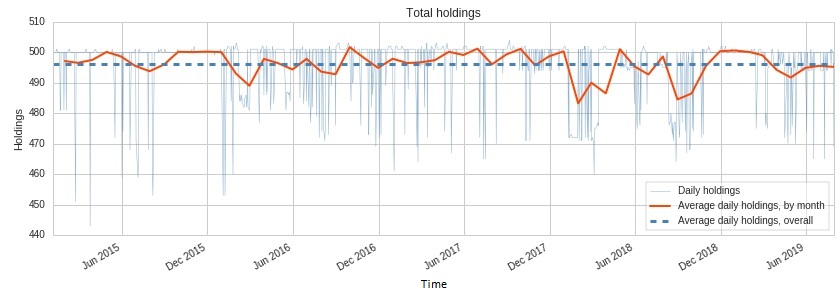}
\caption{Daily Holdings}

\label{p15}
\end{figure}

\textbf{Daily Holdings:}
From the figure~\ref{p15}, we see the total daily holdings of our portfolio which never exceeds 500. As we set our maximum holding limit in our portfolio to be 500.

\begin{figure}[H]
\centering
\includegraphics[scale=0.7]{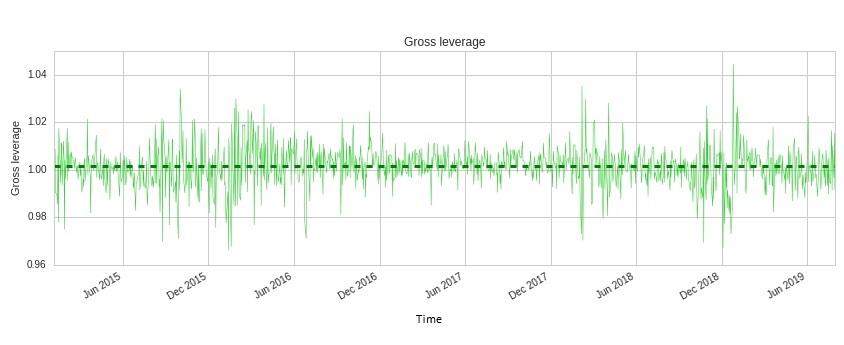}
\caption{Gross Leverage}

\label{p16}
\end{figure}

\textbf{Gross Leverage :} Figure~\ref{p16} shows, we kept our leverage at max 1.05 and at least 0.96 so that our money would be utilized but avoided the risk of being liquidated.

All the features that we calculated were later filtered out and grouped out into their specific dates for trading where they perform the best. The three categories are weekly trading, monthly trading and daily trading. We then used specific different algorithms to trade in order to compare their performance.

\subsection{Time Complexity Analysis of the Model}
In table~\ref{timecom} n is the number of training examples and p is the number of features. 
\begin{table}[H]

\caption{Time-complexity of the proposed methods and Machine Learning Algorithms}
\begin{center}
\begin{tabular}{|l|l|}
\hline
\rowcolor[HTML]{FFCE93} 
Algorithm Name & Time Complexity \\ \hline
Proposed Feature Selection Method &$O(pn^3)$ \\ \hline
Feature Reduction Using ANOVA & $O(pnlogn)$ \\ \hline
Naive Bayes & $O(np)$\\ \hline
Logistic Regression &  $O(p^2n+p^3)$\\ \hline
Stochastic Gradient Descent &  $O(pn^2)$\\ \hline
Support Vector Machine &  $O(n^2p+n^3)$\\ \hline
AdaBoost & $O(np^2)$ \\ \hline
Random Forest & $O(n^2pn\textsubscript{trees})$ \\ \hline
Decision Tree &  $O(n^2p)$ \\ \hline
\end{tabular}
\label{timecom}
\end{center}
\end{table}

The proposed feature selection takes a lot of time as it uses 4 different types of analysis. The other feature reduction algorithm and the other standard machine learning are also mentioned in the table. In the ensemble learning method, each algorithm had to be trained individually before the final result. Therefore the combination of the most efficient algorithm generated the best result as in the Quantopian platform time is an important factor.

\section{Evaluation on Synthetic Dataset}
To further evaluate our model we created 2 synthetic datasets . In order to test if our model works both for normally distributed dataset and non-Gaussian dataset we use 2 different types of data generation techniques. All of the 148 data were used in both the datasets. After running our feature selection model 25 of the features were selected for final decision making.  

\begin{figure}[H]
\centering
\includegraphics[scale=0.45]{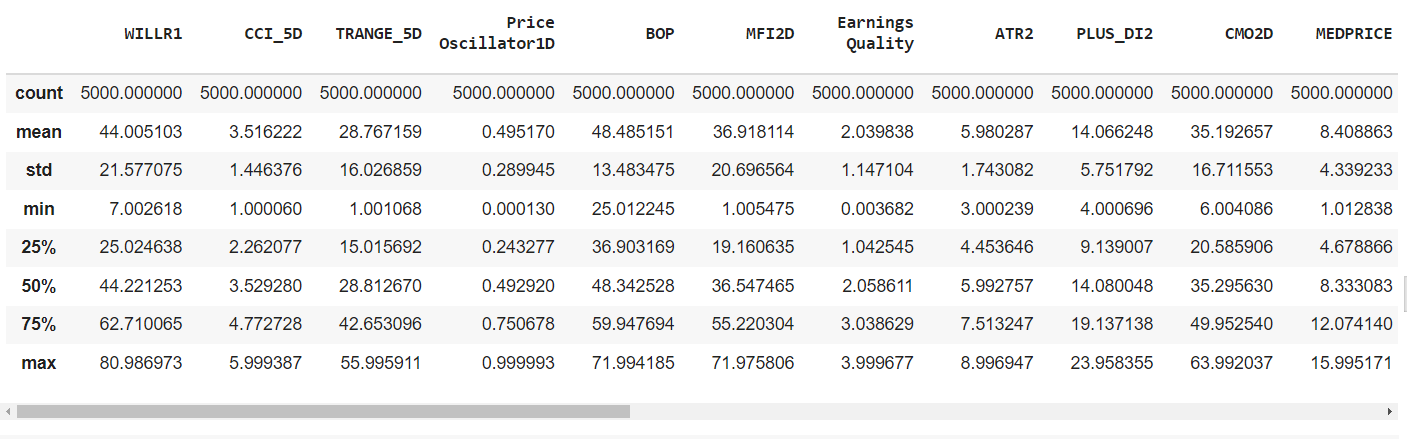}
\caption{ Descriptive Statistics of the Synthetic data}
\label{S0}
\end{figure}
Figure~\ref{S0} shows mean, standard deviation, minimum value, maximum value and specific percentile scores that give a basic idea about the synthetic datasets.

\begin{figure}[H]
\centering
\includegraphics[scale=0.5]{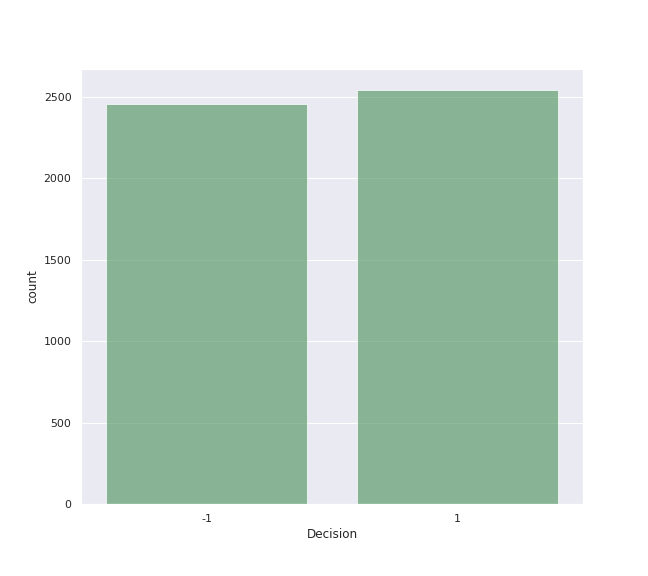}
\caption{Labels of Dataset 1 (Uniform)}
\label{S1}
\end{figure}

In figure~\ref{S1}, the labels are for long and short are represented by 1 and -1 respectively. The number of each label is close to 2500 adding up to a total of 5000 instances.

\begin{figure}[H]
\centering
\includegraphics[scale=0.9]{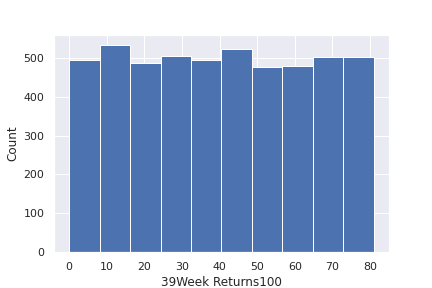}
\caption{Distribution of features in Dataset 1 (Uniform)}
\label{S2}
\end{figure}

In figure~\ref{S2} the distribution of one of the features of the dataset 1 is shown. The min and max value of the data was required to generate this non-Gaussian distribution.  

\begin{figure}[H]
\centering
\includegraphics[scale=0.31]{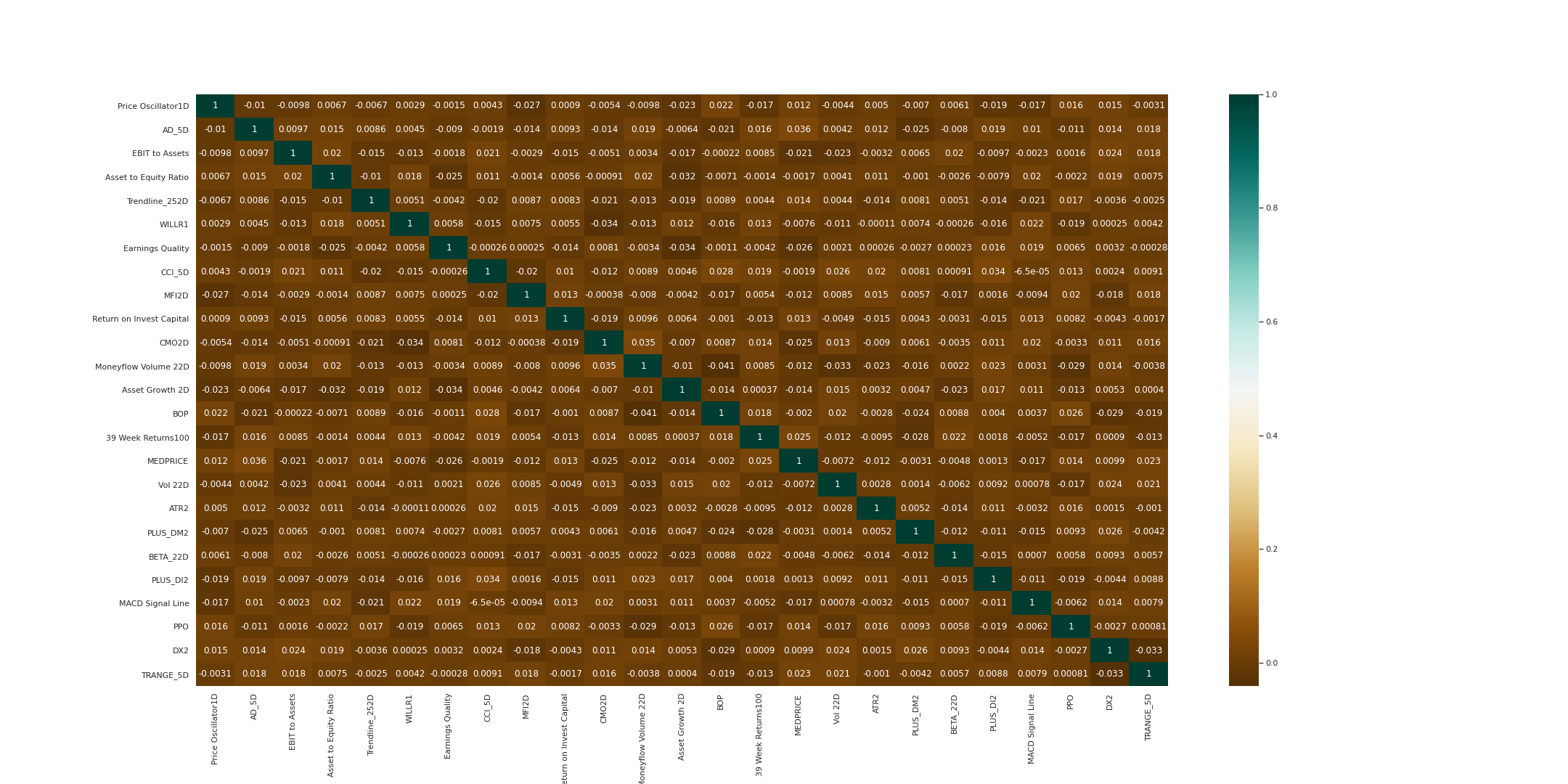}
\caption{Correlation heatmap of Dataset 1 (Uniform)}
\label{S3}
\end{figure}

The correlation of the final 25 features is shown in figure~\ref{S3}. The correlation of the features is important in determining the relationship between the features. Only one feature should out of two if they are highly co-related.

\begin{figure}[H]
\centering
\includegraphics[scale=0.6]{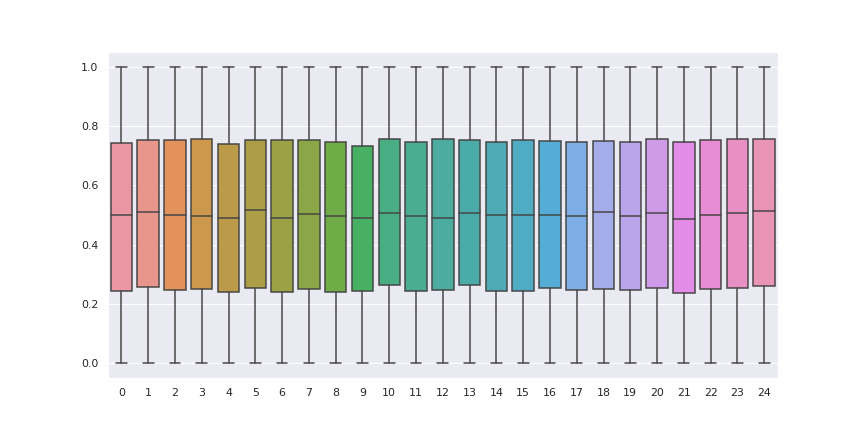}
\caption{BoxPlot of the 25 selected features in Dataset 1 (Uniform)}
\label{S4}
\end{figure}
Figure~\ref{S4} shows the boxplot of the 25 selected features. The values are scaled from 0 to 1. The middle line shows the mean of the feature. The distribution of the feature values can be visualized from the figure~\ref{S4}.

\begin{figure}[H]
\centering
\includegraphics[scale=0.7]{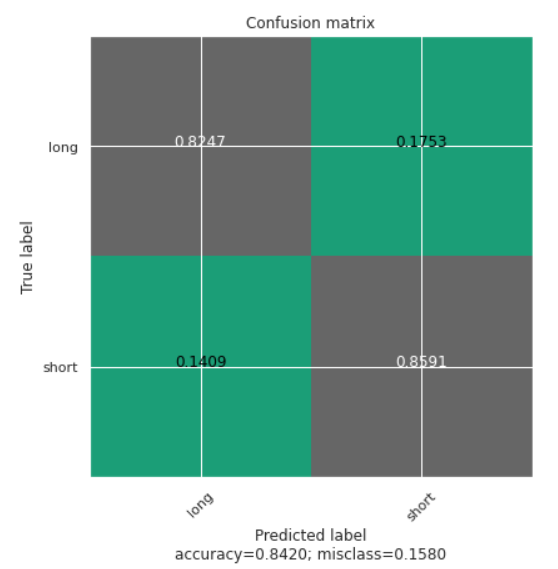}
\caption{Confusion Matrix from the ensemble method on Dataset 1 (Uniform)}
\label{S5}
\end{figure}

Figure~\ref{S5} is the confusion matrix generated by the proposed best model. Our model achieves 84.20\% accuracy in the non-Gaussian dataset. Using The values from the confusion matrix the Recall is 82.46\%, Precision is 85.26\% and the F1 score is 83.84\%.

\begin{figure}[H]
\centering
\includegraphics[scale=0.5]{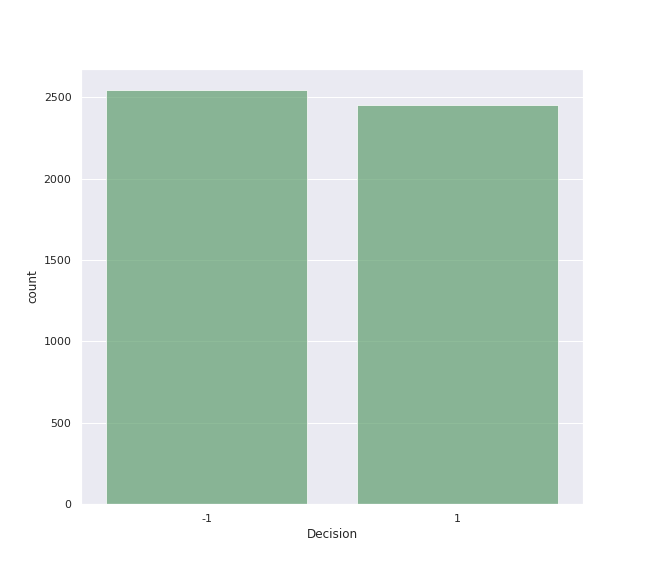}
\caption{Labels of Dataset 2 (Gaussian)}
\label{L1}
\end{figure}
In figure~\ref{L1}, the labels are for long and short are represented by 1 and -1 respectively. The number of each label is close to 2500 adding up to a total of 5000 instances.

\begin{figure}[H]
\centering
\includegraphics[scale=0.9]{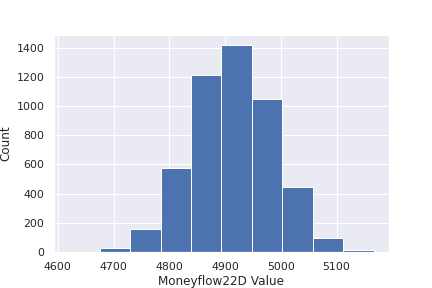}
\caption{Distribution of features in Dataset 2 (Gaussian)}
\label{L2}
\end{figure}

In figure~\ref{L2} the distribution of one of the features of the dataset 2 is shown. The dataset is created using the mean value and standard deviation of the original data.

\begin{figure}[H]
\centering
\includegraphics[scale=0.31]{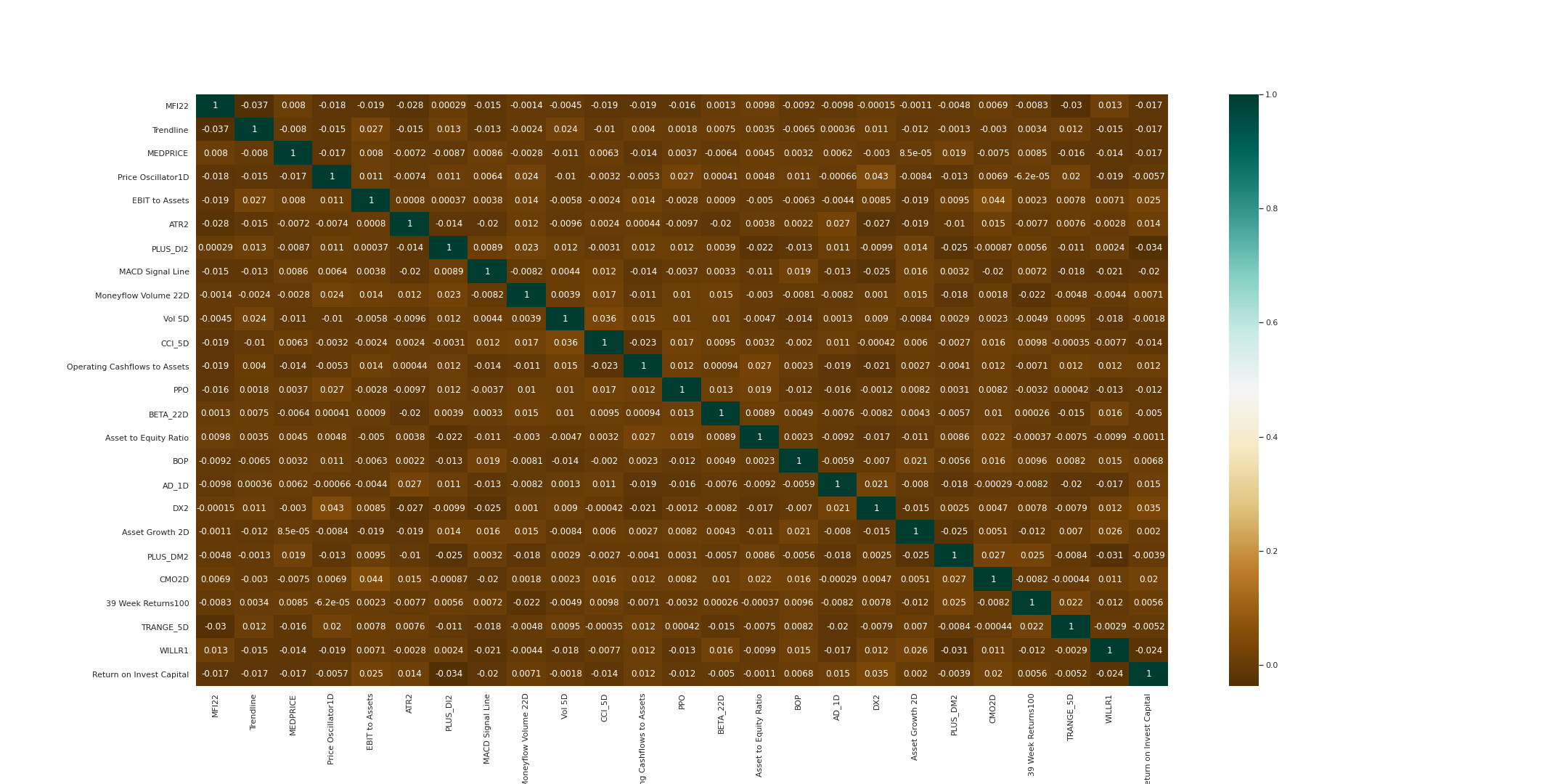}
\caption{Correlation heatmap of Dataset 2 (Gaussian)}
\label{L3}
\end{figure}
The correlation of the final 25 features is shown in figure~\ref{L3}. The correlation of the features is important in determining the relationship between the features. Only one feature should out of two if they are highly co-related.

\begin{figure}[H]
\centering
\includegraphics[scale=0.6]{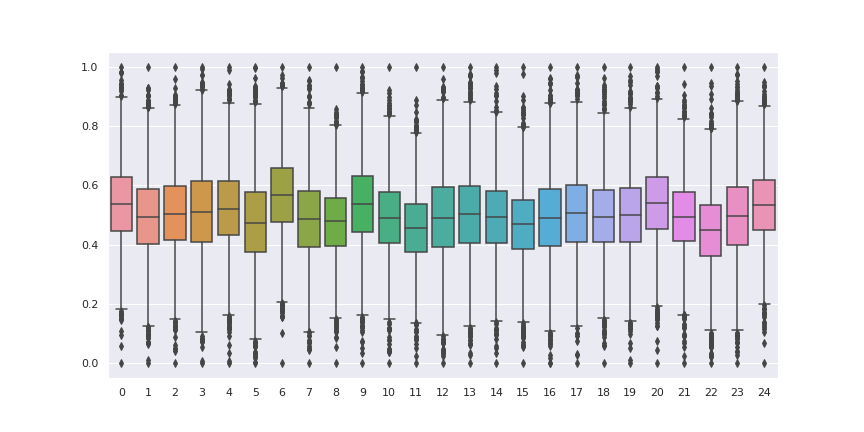}
\caption{BoxPlot of the 25 selected features in Dataset 2 (Gaussian)}
\label{L4}
\end{figure}
Figure~\ref{L4} shows the boxplot of the 25 selected features. The values are scaled from 0 to 1. The middle line shows the mean of the feature. The distribution of the feature values can be visualized from the figure~\ref{S4}.

\begin{figure}[H]
\centering
\includegraphics[scale=0.7]{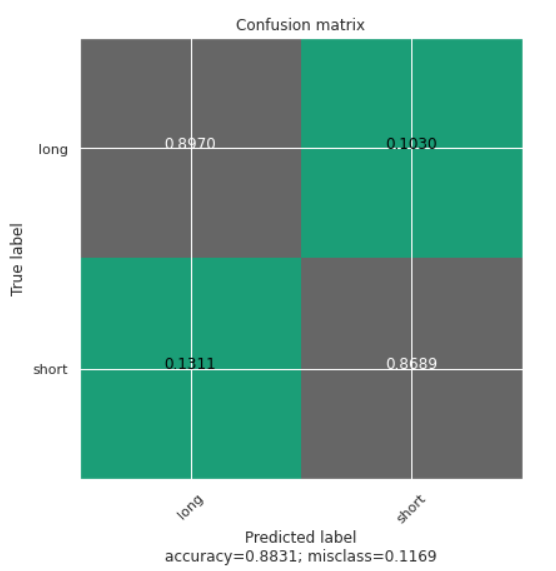}
\caption{Confusion Matrix from the ensemble method on Dataset 2 (Gaussian)}
\label{L5}
\end{figure}
Figure~\ref{L5} is the confusion matrix generated by the proposed best model. Our model achieves 88.30\% accuracy in the non-Gaussian dataset. Using The values from the confusion matrix the Recall is 89.70\%, Precision is 87.36\% and the F1 score is 88.51\%.

\begin{figure}[H]
\centering
\includegraphics[scale=0.7]{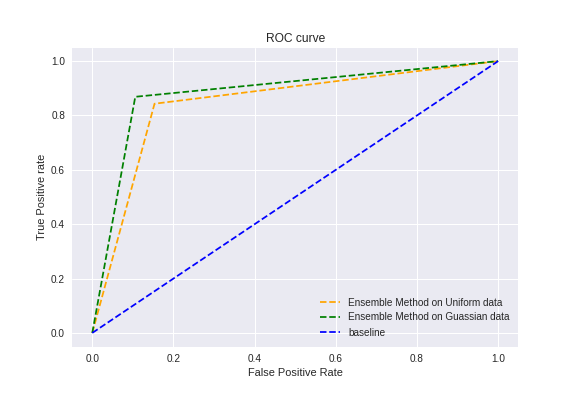}
\caption{ROC curve of the Synthetic Datasets}
\label{L6}
\end{figure}
A greater X-axis value in a ROC curve implies a larger number of False positives than True negatives. While a higher Y-axis value implies a greater number of True positives than False negatives, a lower Y-axis value suggests a lower number of True positives. As a result, the threshold is determined by the capacity to balance False positives and False negatives.

It is evident from the figure~\ref{L6} that the AUC for the ensemble model on Gaussian data ROC curve is higher than that for the non-Gaussian data ROC curve. Therefore, we can say that our model does a better job of classifying the positive class in the normally distributed dataset. The AUC value of our model in the normally distributed dataset is 0.8814 and the AUC value of our model with uniformly distributed dataset is 0.8449.

\begin{table}[H]
\caption{Proposed model performance on Synthetic data}
\begin{tabular}{|l|l|l|l|l|l|}
\hline
\rowcolor[HTML]{FFCE93} 
Type of The Dataset & Accuracy & Precision & Recall & F1-score & AUC-score \\ \hline
Normally Distributed & 88.30\% & 87.36\% & 89.70\% & 88.51\% & 0.8814 \\ \hline
Uniformly Distributed & 84.20\% & 85.26\% & 82.46\% & 83.84\% & 0.8449 \\ \hline
\end{tabular}

\label{perfsyn}
\end{table}

Table~\ref{perfsyn} shows that the proposed model works better when dataset is normally distributed. As most stock data is normally distributed that is why the proposed model is finely tuned to work better with normally distributed data. However,the model also performs quite well achieving 84.20\% accuracy and quite good in other performance matrices as shown in table~\ref{perfsyn}.

\chapter{Discussion and Conclusion }

\section{Comparison with other Models}
The most important part of our model is our novel feature calculation and selection method. For this reason even with the huge drawbacks of the Quantopian platform our model performs on par with the state-of-the-art models that perform quantitative Trading as can be seen from table~\ref{compare}. The biggest advantage of our proposed model is that the feature selection method can be added to any decision making model to make better predictions. 

\begin{table}[H]
\caption{Comparison of Proposed model with state-of-the-art models}
\begin{tabular}{|p{0.2\linewidth}|p{0.15\linewidth}|p{0.1\linewidth}|p{0.15\linewidth}|p{0.25\linewidth}|}
\hline
\rowcolor[HTML]{FFCE93} 
Author & Time Period & Returns & Trading Frequency & Method Used \\ \hline
T. Dai, A. Shah and H. Zhong (2012) & 5 years & 30.66\% & 30 Days & SVM and Logistic Regression \\ \hline
G. Chen, Y. Chen and Fushimi (2017) & 7 years & 103\% & 30 Days & LSTM \\ \hline
Vo, N. N. Y., He, X., Liu, S., \& Xu, G. (2019) & 3 years & 50.78\% & 1 year & Reinforcement Learning \\ \hline
Proposed Model & 8 years & 54.35\% & 1 Day & Ensemble Learning with Feature Selection \\ \hline
\end{tabular}

\label{compare}
\end{table}

Table~\ref{compare} shows that the proposed model performs better than the model of T. Dai and G. Chen. Moreover, the feature extraction and selection method can select the best feature for trading in any time-period. Therefore, this model can be incorporated with any model to significantly improve the quality of the features.

\section{Findings and Research Challenges}

Through experimentation, it is clear that ensemble learning produced a better result in case of stock market trading as compared to using a single algorithm. Furthermore, it also became clear that most important part of a stock trading algorithm is the feature extraction part. The 1 day trading algorithm made 54.35\% over the course of 8 years profit due to the quality of the features that were used for 1 day trading. Whereas the weekly and the monthly algorithm did not perform that as well due to its features. Our most significant contribution is that we detected using statistical measures that which features should work well for which time-frame. The model can clearly capture the trend of the market over a one day period.

The Quantopian platform does not allow users to download their data; therefore, the model was restricted to the limitations of Quantopian. The highest data look back for daily trading was 200 days before the current trading day. In the case of weekly and monthly trading, the days were 150 days and 100 days, respectively. Due to these constraints, the weekly and the monthly trading algorithm could not perform as well as the daily trading algorithm.

Due to our limited resources, we were not able to use Pipeline to train our model and as such, were not able to train our models without exceeding the time limit for some algorithms. It is also difficult to implement high frequency trading such as hourly trading, as Quantopian does not provide features for hourly trading. We were also unable to implement any kind of neural networks as Quantopian would not allow us to import Keras for tensor flow. Furthermore, if we had better access to trading data, we would have been able to run our own neural network over the data for better results, but were unable to do so as Quantopian does not allow downloading of its datasets. With better resources, and better access to trading data, we would have been able to produce better and more accurate results. 

\section{Conclusion and Future works}
This thesis applied novel methods on the stock market of the USA and validated the data in an external synthetic dataset. In this research, we demonstrated a new feature selection method for trading with different time-frames. The results reflect success of the model in a live trading environment.
For future implementation purposes, we intend to design our own reinforcement learning algorithm that will be specifically tailored for this purpose. In order to get better results, we would like to try high-frequency trading, preferably minutely and hourly. There is also a new platform called QuantConnect which offers more flexibility than Quantopain where we can do our future work without the stated limitations. 

\section{Information Sharing Statement}
In order to ensure reproducibility of our research we published our entire work at: \newline{\url{https://github.com/amanat9/QuantopianThesis}}.

\phantomsection
\printbibliography 
\addcontentsline{toc}{chapter}{Bibliography}

\begin{appendices} 

\newpage
\phantomsection
\addcontentsline{toc}{chapter}{Appendix A: Feature Formulas}

\chapter*{Appendix A: Feature Formulas}

\section*{Primary Features} 
Our primary data is the daily (1 day) open, low, high, close, volume of each stock in our tradable universe. Additionally we are retrieving daily balance sheet, cash flow statement, income statement, operation ratios, earning report, valuation, valuation ratios are taken as primary data. Using these primary data the 4* secondary factors were created. These factors are commonly used in financial prediction by traders.\newline
bs = morningstar.balanceSheet \newline
cfs = morningstar.cashFlowStatement \newline
is = morningstar.incomeStatement \newline
or = morningstar.operationRatios \newline
er = morningstar.earningsReport \newline
v = morningstar.valuation \newline
vr = morningstar.valuationRatios \newline

\section*{Secondary Features}
\textbf{Balance\_sheet :-}
Balance sheet is a very important financial statement and is both financial modelling and accounting. It is used to portray a company's total assets. The balance sheet is usually calculated using the equation as follows.

\begin{align}
\begin{array}{l}
A =  L - SE\\
where:\\
A = Assets\\ L = Liabilities\\ SE = Shareholders\,Equity\\
\end{array}
\end{align}

\textbf{Cash Flow Statement( CFS )}
It is a measure of how a company manages its financial strength and liquidity. It has a very high correlation ship with balance sheet and income statement. It can be used to analyze a company.

\textbf{Income Statement:}
It is used for reporting a company's financial status over a specific accounting period. It summarizes a company's total returns, expenses over a period of time.
\begin{align}
\begin{array}{l}
Net\,Income = \mathrm{R}+\mathrm{G}-\mathrm{E}+\mathrm{LE}\\
where:\\
\mathrm{R} =  revenue\\
\mathrm{G} =  gains\\
E = expenses \\\mathrm{LE} =  losses\,equity
\end{array}
\end{align}

\textbf{Operating Ratio:}
It shows the economy of a company by comparing total operating expenses to company net sales. The less the ratio the more efficient the company is at generating revenue. 

\begin{align}
\begin{array}{l}
Operating Ratio =\frac{O E+C G}{\text {Net sales}}\\
where :\\
\mathrm{OE}= Operating\,expenses\\ \mathrm{CG}= cost\,of\,goods\,sold
\end{array}
\end{align}

\textbf{Earnings Report:}
It is a quarterly earnings report made by companies to report their companies.

Valuation:-
It is used to determine the current worth of assets of a company.

\textbf{ADX and DX:} It is a technical index used to indicate the strength of the trade. This strength can either be positive or negative and this is shown by two indicators +DI and -DI thus ADX commonly includes 3 separate lines. Additionally, It is a technical indicator that is used to predict the divergence side of the market. The two components of DMI are +DI and -DI.
\begin{align}
\begin{array}{l}{\mathrm{DI_{Plus}}=\left(\frac{\text { Smoothed }+\mathrm{DM}}{\mathrm{ATR}}\right) \times 100} \\ {\mathrm{DI_{Minus}}=\left(\frac{\text { Smoothed }-\mathrm{DM}}{\mathrm{ATR}}\right) \times 100} \\ {\mathrm{DX}=\left(\frac{|\mathrm{DI_{Plus}}-\mathrm{DI_{Minus}}|}{|\mathrm{DI_{Plus}}+\mathrm{DI_{Minus}}|}\right) \times 100} \\ {\mathrm{ADX}=\frac{(\text { Prior } \mathrm{ADX} \times 13)+\text { Current } \mathrm{ADX}}{14}}\end{array}
\end{align}

\textbf{APO:}
It finds the absolute value and finds the difference between two different exponential moving averages . When the APO indicator goes above zero we go long i.e. Bullish and below zero we go short i.e. bearish.

\begin{align}
\begin{array}{l}
A P O= FEMA -  SEA\\
where:\\
FEMA=Fast\,Exponential\,Moving\,Average\\
SEA=Slow\,Exponential\,Average
\end{array}
\end{align}

\textbf{Mean Revision Theory:}
It is used to statistically analyze the market condition, which can overall affect the trading strategy. Mean revision also takes advantage of extreme price fluctuation of particular stocks. They can be applied for both buying and selling strategies.

\begin{align}
\begin{array}{l}
Mean\,revision =(\mathrm{MR}-\mathrm{Mean}(\mathrm{MR}))-\mathrm{Std}(\mathrm{MR})\\
where:\\
\mathrm{MR}= Monthly\,returns\\Std= Standard\,deviation
\end{array}
\end{align}

\textbf{CMO:}
It is very similar to other similar momentum oscillators. It calculates momentum for both Market Up and Down days but it does not smooth out the results. The oscillator indicates between +100 and -100.

\begin{align}
\begin{array}{l}{\text { Chande Momentum Oscillator }=\frac{s H-s L}{s H+s L} \times 100} \\ {\text { where: }} \\ {s H=\text { high close summation in } \mathrm{N} \text { periods }} \\ {s L=\text { low close summation in } \mathrm{N} \text { periods }}\end{array}
\end{align}

\textbf{Returns}

It is an indication of total money made or lost during transactions. Returns can be expressed as a ratio of profit to investment.

\begin{align}
\text {Rate of Returns}=\frac{\text {current value-initial value}}{\text {initial value}} \times 100
\end{align}

\textbf{Williams \%R}
It is known as Williams perfect Range, which is a type of momentum calculator that has an indicator range between 0 to -100 and measures the level of overbought and oversold. It is used to find the most optimal time for entry and exit the market.

\begin{align}
\begin{array}{l}{\text { Wiliams } \% R=\frac{\text { Highest High }-\text { Close }}{\text { Highest High }-\text { Lowest Low }}} \\ {\text { where }} \\ {\text { Highest High }=\text { Peak price in the lookback }} \\ {\text { time period, typically } 2 \text { weeks. }} \\ {\text { Close = Latest closing price. }} \\ {\text { Lowest Low = trough level price in the lookback }} \\ {\text { time period, typically 2 weeks. }}\end{array}
\end{align}

\textbf{ATR:}
It measures the market volatility, by dissolving the entire range of an asset price. Stocks with higher volatility has higher ATR and vice versa. This acts as an indicator for traders to exit and enter a trade.

\begin{align}
\begin{array}{l}T Y=\max \left[(H-L),\left|H-C_{p r e v}\right|,\left|L-C_{p r e v}\right|\right]\\ {A T R=\frac{1}{n} \sum_{i=1}^{n} T R_{i}} \\ \\ {\text 
{\textbf{where:}}} \\ {H=\text { High }}\\{L=\text { Low }}\\{C=\text { Close }}\\{T R_{i}=\text { a particular true range; and }} \\ {n=\text { the time period employed (usually } 14 \text { days) }}
\end{array}
\end{align}

\textbf{AD:}
This is an indicator that makes use of volume and price to determine if a stock is accumulated or distributed. This factor looks for changes between stock price and volume flow, thus providing a hint of how strong a trend is.
\newline
The Formula for the Accumulation / Distribution Indicator.

\textbf{where:}

\begin{align}
\begin{aligned} \mathrm{CMFV} &=\frac{\left(P_{C}-P_{L}\right)-\left(P_{H}-P_{C}\right)}{P_{H}-P_{L}} \times \mathrm{V} \\ CMFV &= \text { Current Money Flow Volume} \\ P_{L} &= \text { Losing price } \\ P_{L} &=\text { Low price for the period } \\ P_{H} &=\text { High price for the period } \\ \mathrm{V} &=\text { Volume for the period } \end{aligned}
\end{align}

\textbf{BETA:}
A coefficient measure of volatility for an individual stock in contrast to the entire market. Statistically beta is the gradient of the line. By default the market beta is 1.0. 

\begin{align}
\begin{array}{l}{\text { Beta coefficient }(\beta)=\frac{\text { Covariance }\left(R_{e}, R_{m}\right)}{\operatorname{Variance}\left(R_{m}\right)}} \\ {\text { where: }} \\ {R_{e}=\text { revenue from a stock }} \\ {R_{m}=\text { revenue from overall market }} \\ {\text { Covariance = Correlation of returns of a stock  to }} \\ {\text {  returns of the market  }} \\ {\text { Variance = Divergence of the market's value }} \\ {\text {from average }}\end{array}
\end{align}

\textbf{MP:}
It calculates the mean of the high and low of a stock candle.
\begin{align}
MedPrice=(h i g h(t)+\operatorname{low}(t)) / 2
\end{align}

\textbf{MFI:}
It is a technical indicator that makes use of price and volume as a reference to identify if a stock is overvalued or undervalued. It can be used to spot the change in the daily price of the stock.
The value of the oscillator ranges between 0-100.

\begin{align}
\begin{array}{l}{\text { Money Flow Index }=100-\frac{100}{1+\text { Money Flow Ratio }}} \\ {\text { where: }} \\ {\text { Money Flow Ratio }=\frac{14 \text { Period Positive Money Flow }}{14 \text { Period Negative Money Flow }}} \\ {\text { Raw Money Flow }=\text { Common Price } * \text { Volume }} \\ {\text { Common Price }=\frac{(\text { High }+\text { Low }+\text { Close })}{3}}\end{array}
\end{align}

\textbf{PPO:}
This is used to show the correlation ship between two Moving averages between 0-1. It compares asset benchmarks, market volatility to come up with trend signals and help predict the trend of the market.

\begin{align}
\begin{array}{l}{\mathrm{PPO}=\frac{12\text {period} \mathrm{EMA}-26\mathrm{period } \mathrm{EMA}}{26\mathrm{period } \mathrm{EMA}} \times 100} \\ {\text { Signal Line }=9 \text { -period EMA of PPO }} \\ {\text { PPO Histogram }=\mathrm{PPO}-\text { Signal Line }}
\\ \\
\textbf{Where:}
\\
\text {EMA = Exponential Moving Average}
\end{array}
\end{align}

\textbf{Asset to Equity Ratio:}
It shows the correlation between the assets owned by a firm to the total percentage of the shareholders. The higher the ratio the greater the firm’s debt.

\textbf{Capex to Cash Flow:}
This is used to estimate a company's long term assets and also how much cash a company is able to generate.

\begin{align}
\begin{array}{l}
\text {Cash to capital Expenditures}=\frac{\text {Cash Flow from Operation}}{\text {Capital Expenditure}}
\end{array}
\end{align}

\textbf{Asset Growth:}
It is the growth of the overall asset of a company.
\begin{align}
\begin{array}{l}
\text {Asset Growth}=\frac{\text {Asset value prior }-\text { Asset value current}}{\text {Asset value prior}} \times 100
\end{array}
\end{align}

\textbf{EBIT to Asset:}
It is a sign of a company's benefits which is generated from operations and trades and ignores tax burden and capital structure.

\begin{align}
\begin{array}{l} {\text { EBIT = R - COGS - OE }} \\ {\text { Or }} \\ {\text { EBIT = NI + I + T }} \\ \\ {\text { \textbf{where}: }} \\{\text { R = Revenue }}\\ {\text { NI = Net Income }}\\ {\text { I = Interest }}\\{\text { T = Taxes }}\\{\text { OE = Operating Expenses }}\\ {\text { COGS = Cost of goods sold }}\end{array}
\end{align}

\textbf{EBITDA Yield:}
It is usually reported as a Quarterly earnings press release. It ignores taxes and non-operating expenses thus highlighting only important for the market analyst to focus on.

\textbf{MACD Signal Line:}
This indicator shows the relationship between different stock's moving averages. After calculation, a nine day EMA-line is drawn over MACD line to use as a buy or sell indicator.
\begin{align}
\mathrm{MACD}=12\mathrm{period} \mathrm{EMA}-26\mathrm{period} \mathrm{EMA}
\end{align}

\textbf{Money Flow Volume:}
Money flow is an indicator of when and at which price the stock was purchased. If more security was bought during the uptick time compared to downtick then the downtick time then the indicator is positive because almost all the investors participating in the trade were willing to give a high price for the stock and vice versa.

\textbf{Operating Cash Flows to Assets:}
It is the flock of revenue generated by a company's normal business operation. It is an indicator of whether a company can generate a substantial amount of positive cash flow to maintain its growth. This indicator gives the market analysts a clear view of what a company is capable of.

\textbf{Return on Invest Capital:}
 This gives the market analyst an indicator of how well a company uses its resources to generate its revenue.
\begin{align}
\begin{array}{l}{\mathrm{ROIC}=\frac{\mathrm{NOPAT}}{\text { Invested Capital }}} \\ {\text { where: }} \\ {\text { NOPAT = Net operating profit after tax }}\end{array}
\end{align}

\textbf{39 Week Returns:}
Gives us the total returns over a period of 39 weeks
\begin{align}
\begin{array}{l}
39 Week Returns =\frac{R(T)-R(T-215)}{R(T)} \times 100\\ where:\\
\mathrm{R}= Returns
\end{array}
\end{align}

\textbf{Trend line:}
It shows the momentum / Trend of the market from one given point to another. \\
\textbf{Volume 22 Days:}
It gives us the total amount of volume generated over a period of 22 days.

\newpage
\phantomsection
\addcontentsline{toc}{chapter}{Appendix B: Machine Learning Algorithms}

\chapter*{Appendix B : Machine Learning Algorithms }


\section*{Naive Bayes }

Naive Bayes classification uses Bayes' rule. Assuming 
Ck= a particular event 
x= feature vector 
\begin{align}
P\left(c_{k} | \mathbf{x}\right)=P\left(c_{k}\right) \times \frac{P\left(\mathbf{x} | c_{k}\right)}{P(\mathbf{x})}
\end{align}

Bayes' rule defines  the probability of  a particular event ck occurring for feature vector x, can be computed from the given formula.

For estimating $P(c_k|x)$ from a dataset we must first compute  $P(c_k|x)$. The strategy used to find the distribution of x conditional on ck is specified by the following formula:
\begin{align}
P\left(\mathbf{x} | c_{k}\right)=\prod_{j=1}^{d} P\left(x_{j} | c_{k}\right)
\end{align}

In this formula, we assume that $x_j$ having a particular value is independent of the occurrence of any other $x_j$’ from the x feature vector for a particular event $c_k$. By plugging in the estimates the equation 2 becomes:

Gaussian Naive Bayes is better for this case as the features are continuous. Whereas Bernoulli Naive Bayes works better when the features are binary.

 \section*{Logistic Regression  }
       
Regression models are widely used for data-driven decision making. In many fields, logistic regression has become the standard method of data analysis in such situations. The key to any such kind of analysis is to find the model that fits best when explaining the relationship between a dependent and one or more independent variables. Unlike linear regression which most people are familiar with, where the outcome variable is usually continuous, a condition for logistic regression is that the outcome variable is binary. Logistic regression also allows us to determine to what degree a chosen independent variable affects the outcome.

Two reasons why logistic regression is so widely used is that it is 1) flexible and can be easily used in many situations, and 2) it allows for meaningful interpretations of the results. For simplicity, the quantity $\gamma(x)=E(Y given\,x)$ is used to represent Y's conditional mean, given a value x.

The logit transformation is integral for logistic 
regression. The transformation is as follows:

\begin{align}
\begin{aligned} p(x) &=\ln \left[\frac{\gamma(x)}{1-\gamma(x)}\right] \\ &=\theta_{0}+\theta_{1} x \end{aligned}
\end{align}

The importance of the logit, p(x) lies in the fact that it contains many useful properties of 
Logistic regression. The logit takes linear vales which might be continuous, either positive or negative and depends on x range.

To summarize, when the outcome variable is dichotomous, in regression analysis:

\begin{itemize}
  \item The logistic regression mean must be scaled to be between 1 and 0. 
  \item The binomial, as opposed to the normal distribution, describes the distribution of errors.
   \item The principles of linear regression can also be applied to logistic regression.

\end{itemize}

For the model to produce reliable results, we need to have a large number of observations (at least 50).

In order to prevent overfitting of data, L1 (Lasso) and L2 (Ridge) regression is used. The difference between these two methods of regularization lie in the penalty term. L2 uses the “squared magnitude” as the penalty term to the loss function, while L2 uses the “absolute value of magnitude”.

\begin{align}
\begin{array}{c}{\sum_{i=1}^{n}\left(y_{i}-\sum_{j=1}^{p} x_{i j} \alpha_{j}\right)^{2}+\gamma \sum_{j=1}^{p} \alpha_{j}^{2}} \\ {\text { cost function }}\end{array}
\end{align}

\begin{align}
\begin{array}{c}{\sum_{i=1}^{n}\left(y_{i}-\sum_{j=1}^{p} x_{i j} \alpha_{j}\right)^{2}+\gamma \sum_{j=1}^{p}\left|\alpha_{j}\right|} \\ {\text { Cost function }}\end{array}
\end{align}

The above equations show the cost function when using L2 and L1 regularization respectively where $\alpha$ is the weight put on a particular feature x and $\gamma$ is the coefficient of the penalty term.

 \section*{Stochastic Gradient Descent }
It is a first-order optimizing supervised machine learning algorithm that specializes to fit a straight line over a series of data points with the least amount of error.

\begin{align}
w_{t+1}=w_{t}-\gamma \frac{1}{n} \sum_{i=1}^{n} \nabla_{w} Q\left(z_{i}, w_{t}\right)
\end{align}

$w$ is the weight we want to optimize with regards to cost $\nabla Q(z,w)$ where $\gamma$ is the learning rate. It works by first assuming randomly a point of intercept to draw a straight line and for each individual weight of the graph we find the predicted Y value i.e. $\hat{y}$ using which we calculate the lingering or remaining value i.e. the difference between real y and $\hat{y}$ and then square the residual and square its value, and then find the sum of the squared residual for each and every Y value that exists for the X value. If we keep on increasing the intercept and for every intercept we get a sum of squared residual value, which if we plot a graph we will obtain a graph that looks similar to y = $X^2$. The maximum value of the graph will be the one that has the lowest squared residual. This is a very slow method but gradient descent uses this concept but works in a much faster way by taking big steps when it is far away from the optimal value and gradually decreases the step size when it gets closer. Gradient descent derives the sum of the squared residuals with respect to the intercept giving us the slope of the curve at that state of time. The closer we get to intercept the closer the slope gets to 0. We then calculate the step size we multiply the slope that we got with $\alpha$i.e. learning rate. Using the new intercept we got we then repeat the entire working process until we get a slope close to 0 or we reach our iterative limit, when we stop our algorithm.

Stochastic gradient descent on the other hand works very similarly but is very optimized and efficient when it comes to using it on large data sets.\cite{bottou2010large}

What stochastic gradient descent does is it picks random values from the weight and only uses that value to perform the entire working process and so on. Thus this reduces the calculation factor by F-1, here F is sum of the points. It also performs well when using it on data with a lot of redundancy, as it clusters them and only picks a random value from every single cluster to perform the working steps. Thus if there are 5 clusters stochastic gradient descent will pick 5 points to work with. 

Thus stochastic gradient descent works well with stock prediction all whilst trading real time is it reduces the complexity of the algorithm by a whole lot thus not resulting in any TLE unlike gradient descent.

\textbf{Hinge Loss : }It is a loss function that is mainly used to train classifiers using the maximum margin classification.\cite{rosasco2004loss}

\textbf{Elastic Net Penalty :} Elastic net penalty that is mainly used to overcome the limitations Lasso regression. If there are highly correlated values lasso regression usually tends to pick one variable from the group that are highly correlated and ignore the rest but what elastic net does is it adds a squared penalty to it. Adding this term gives this loss function a unique minimum( strongly convex ).

\section*{Support Vector Machine (SVM)}
SVM is a classification and regression based algorithm. It is used to maximize predictive accuracy whilst avoiding the overfitting of data. It is used for applications such as handwriting,  face,  text and hypertext classification, Bioinformatics etc. SVM is used to achieve maximum separation between data points. Hyperplane is a part of SVM that maximize the separation of data points by increasing the line width with increments. It starts by drawing a line and two equidistant parallel lines. Next the algorithm picks a stopping point so that the algorithm does not run into an infinite loop and also picks an expanding factor close to 1 example 0.99. \cite{jakkula2006tutorial}

\section*{AdaBoost} 
A boosting algorithm increases the accuracy of weak learners. A weak learner as an algorithm that uses a simple assumption to output a hypothesis that comes from an easily learnable  hypothesis class and performs at least slightly better than a random guess. If each weak learner is properly implemented then Boosting aggregates the weak hypotheses to provide a better predictor which will perform well on hard to learn problems.

Adaboost is the short form of Adaptive boosting. The AdaBoost algorithm outputs a “strong” function that is a weighted sum of weak classifiers. The algorithm follows an iterative process where in each iteration the algorithm focuses on the samples where the previous hypothesis gave incorrect answers. The weak learner is returns a weak function whose error is et such that 

\begin{align}
\epsilon_{I} \stackrel{\mathrm{def}}{=} L_{\mathbf{D}^{(t)}}\left(h_{t}\right) \stackrel{\mathrm{def}}{=} \sum_{i=1}^{m} D_{i}^{(t)} \mathbb{I}_{\left[h_{t}\left(\mathbf{x}_{i}\right) \neq \mathbf{y}_{i}\right]}
\end{align}

where $L_D$ is the loss function and $h$ is the hypothesis and then a specific classifier is assigned a weight for $h_t$ as follows: $w_{t}=\frac{1}{2} \log \left(\frac{1}{\epsilon_{t}}-1\right)$. So, the weight given to that weak classifier is inversely proportional to the error of that weak classifier. \cite{shalev2014understanding}

\section*{Random Forest}

A random forest uses a set of decision trees. As a class of decision trees of unspecified size has infinite VC dimension (Vapnik–Chervonenkis dimension), we restrict the size of the decision tree in order to prevent overfitting. Creating an ensemble of trees is another way to reduce the probability of overfitting \cite{shalev2014understanding}.

An advantage of using Random Forest is that it both a classifier and regressor \cite{liaw2002classification}. For our purposes, we applied Random Forest for classification. The algorithm works as follows:

Create a bootstrapped dataset from the original data (bagging). An important point about bootstrap samples is that the same sample can be chosen more than once, given they are chosen at random.  
Next, we create decision trees for each sample in our bootstrap dataset. At every tree node, choose a random subset of variables and the best split among those are chosen.
We use the aggregate of the predictions of our trees in order to predict the classification of new data. For classification purposes, we use the majority vote. The average is used for regression. 

We can then easily estimate the error in our results in the following manner:
We take a sample from our original data, which was not used to create our decision trees. This sample is called an “Out of bag” (OOB) sample. We then try to predict the data of the out of bag sample using the tree we grew by applying bootstrapping.
We then aggregate the predictions of the out of bag samples and calculate the rate of error. This is called the OOB estimate of the error rate.

Given enough trees have been grown, the OOB estimate of error rate is significantly accurate.

\section*{Decision Tree}

Decision Tree Classifiers divides a problem into a collection of subproblems turning a complex problem easier to solve [8]. Using entropy as the criteria of splitting tress is useful when the problem contains numerous classes. The objective used for tree design in our model is to minimize uncertainty in each layer, or put differently increase entropy reduction. Shanon’s entropy, defined as

\begin{align}
H=\sum_{i} p_{i} \log p_{i}
\end{align}

Pi= a prior likelihood of class i 
\begin{align}
   Gain(S, A) &= Entropy(S) - \sum_{\upsilon \epsilon A} \frac{S_{\upsilon}}{S} Entropy(S_{\upsilon})
\end{align}

Entropy is used to find the most gain of a particular factor. And the factor with the most gain is used to make a split. At the terminal nodes a decision is given about the classification.

One advantage of decision tree over other classifiers is that a training sample is tested for subsets of classes and not all classes. This reduces computation and improves performance of the classification task. The decision tree classifier also uses different subsets of the features of the given problem. This makes the classifier perform better than single layer classifiers. Decision tree classifier also overcomes high dimensionality problem as it takes limited factors for each decision tree.

Overlapping is one of the problems of using decision tree classifier. The classifier takes a large amount of time and space when the label class is large. But that is not the case in this system as the classification task is binary. There are a lot of difficulties involved in designing an ideal decision tree. Error may also add up in each level to reduce the accuracy of the model \cite{safavian1991survey}.

\end{appendices}

\end{document}